\documentclass{channel}
\usepackage{titlesec}
\usepackage{amssymb, amsfonts}
\usepackage{lipsum}
\usepackage{pslatex}
\usepackage{epsfig,soul}
\usepackage{caption,subcaption}
\usepackage{booktabs,multirow}
\usepackage{graphicx,tikz,graphics,epsfig}
\usepackage{url,CJKutf8}
\usepackage{color,xcolor}
\usepackage{multirow}
\usepackage{physics,amsmath}
\usepackage[square,sort,comma,numbers]{natbib}
\newcommand{\RomanNumeralCaps}[1]
    {\MakeUppercase{\romannumeral #1}}
\titleformat{\paragraph}
{\normalfont\normalsize\bfseries}{\theparagraph}{1em}{}
\titlespacing*{\paragraph}
{0pt}{3.25ex plus 1ex minus .2ex}{1.5ex plus .2ex}

\title{Minimum-dissipation model for large-eddy simulation in OpenFOAM — A study on Channel Flow, Periodic Hills and Flow over Cylinder}
\author{Jing Sun$^{*}$ Roel Verstappen$^{*}$}
\address{$^{*}$ Computational and Numerical Mathematics--Bernoulli Institute\\
University of Groningen\\
Nijenborgh 9, 9747 AG Groningen, The Netherlands}

\begin{document}
\thispagestyle{empty}

%
%
\section{Abstract}
The minimum-dissipation model is applied to turbulent channel flows up to $Re_\tau = 2000$, flow past a circular cylinder at $Re=3900$, and flow over periodic hills at $Re=10595$. Numerical simulations are performed in OpenFOAM which is based on finite volume methods for discretizing partial differential equations. We use both symmetry-preserving discretizations and standard second-order accurate discretization methods in OpenFOAM on structured meshes. The results are compared to DNS and experimental data. 

The results of channel flow mainly demonstrate the static QR model performs equally well as the dynamic models while reducing the computational cost. The model constant $C=0.024$ gives the most accurate prediction, and the contribution of the sub-grid model decreases with the increase of the mesh resolution and becomes very small (less than 0.2 molecular viscosity) if the fine meshes are used. Furthermore, the QR model is able to predict the mean and rms velocity accurately up to $Re_\tau = 2000$ without a wall damping function. The symmetry-preserving discretization outperforms the standard OpenFOAM discretization at $Re_\tau=1000$. The results for the flow over a cylinder show that mean velocity, drag coefficient, and lift coefficient are in good agreement with the experimental data. The symmetry-preserving scheme with the QR model predicts the best results.
The various comparisons carried out for flows over periodic hills demonstrate the need to use the symmetry-preserving discretization or central difference schemes in OpenFOAM in combination with the minimum dissipation model. The model constant of $C=0.024$ is again the best one.

\section{Introduction}
Turbulent flows are a common phenomenon in various engineering applications, but their simulation via direct numerical simulation (DNS) is expensive and even infeasible for high Reynolds flows. The simplified Reynolds-averaged Navier-Stokes (RANS) model introduces large-scale unsteadiness, which shortens the calculation time, but is less accurate. Large-eddy simulation (LES) is introduced to address these limitations. LES resolves the larger-scale, unsteady, turbulent motions directly while modeling the effect of the smaller scale motions. The model represents the unresolved scale of motion and is therefore called sub-grid model.

Among the existing sub-grid models for LES, the Smagorinsky model is the most commonly used one\cite{smagorinsky1963general}. Although the Smagorinsky model gives satisfactory results in decaying homogeneous isotropic turbulence simulations\cite{lilly1966application}\cite{mansour1979improved}, it inappropriately dissipates eddies for laminar and transitional flows. One way to enhance the performance of the Smagorinsky model is to compute the model constant dynamically, but it is computationally expensive. Another approach is the wall-adapting local eddy-viscosity (WALE) model, which corrects behavior near walls using the square of the velocity gradient tensor. There is also the Vreman model, which is insensitive to pure shear but can yield eddy dissipation for back-scatter and solid body rotation.

Minimum-dissipation models are a simple alternative to the Smagorinsky-type approaches to parametrize the subfilter turbulent fluxes in large-eddy simulation. The first minimum-dissipation eddy-viscosity model is the QR model proposed by Verstappen \cite{verstappen2011does,verstappen2018much}. 
The QR model has many desirable properties. It is more cost-effective than the dynamic Smagorinsky model, it appropriately switches off in laminar and transitional flows, and it is consistent with the exact subfilter stress tensor on isotropic grids. 
Subsequently, the anisotropic minimum-dissipation model (AMD) is developed by Rozema et al. for the flow on anisotropic grids \cite{rozema2015minimum}. Abkar and Moin used the AMD model to study the high-Reynolds-number rough-wall boundary-layer flow \cite{abkar2016minimum}. Zahiri et al. implemented the AMD model into OpenFOAM and tested it on single-phase and multi-phase flows by simulating a low-Reynolds number channel flow, a temporal mixing layer and a flow over a 3D sphere \cite{zahiri2019anisotropic}. Lasota et al. applied the AMD model to hybrid aeroacoustic simulations of human phonation \cite{lasota2023anisotropic}. 
However, few studies have investigated the QR model, especially in open-source software. In this work, we implement the QR model in OpenFOAM and perform simulations in high-Reynolds-number and complex geometries, making this the first study of its kind.

Regarding the numerical errors encountered in computational fluid dynamics (CFD), three primary types are typically identified: round-off, iterative, and discretization errors. Round-off errors arise due to the finite precision of floating-point calculations on computers; however, they are generally considered negligible when utilizing double-precision machines. Iterative errors, stemming from the nonlinearity of governing equations, are typically small once the solution has sufficiently converged. The discretization error, however, is the result of discretizing the governing partial differential equations into algebraic equations and is considered dominant among the numerical errors in CFD simulations \cite{ye2023verification}. Komen et.al reported that the numerical errors in turbulent channel flow at $Re_\tau=180$ result in a net numerical dissipation rate that is larger than the subgrid-scale dissipation rate \cite{KOMEN2017565}. Castiglioni and Domaradzki \cite{castiglioni2015numerical} demonstrated that the numerical dissipation can be significantly larger than the dissipation of the classical Smagorinsky SGS model in an LES of the flow over a NACA 0012 airfoil using a commercial CFD code.

To address the issue of discretization error, researchers have proposed symmetry-preserving discretization techniques. Morinishi et al.\cite{morinishi1998fully} reviewed existing conservative, second-order finite-difference schemes for structured meshes, and introduced a “nearly conservative”  fourth-order scheme. 
Verstappen and Veldman (2003)\cite{verstappen2003symmetry} proposed to exactly preserve the symmetry properties of the underlying differential operators on the unstructured staggered grid. The basic idea behind this approach is mimicking the crucial symmetry properties of the underlying differential operators, i.e., the convective operator is represented by a skew-symmetric matrix and the diffusive operator by a symmetric, positive-definite matrix. Trias et al.\cite{TRIAS2014246} generalized this method for unstructured collocated meshes and proposed an approach, based on a fully-conservative regularization of the convective term,  to mitigate the checkerboard spurious modes.
Building upon these ideas, Komen et al.\cite{komen2021symmetry} developed a conservative symmetry-preserving second-order time-accurate PISO-based pressure-velocity coupling method for solving the incompressible Navier-Stokes equations on unstructured collocated grids. They implemented this approach in OpenFOAM. The code used in the present study is provided by Hopman \cite{janneshopmanRKSymFoam}. 

In this study, we validate the effectiveness of the combined minimum-dissipation model and symmetry-preserving discretization in simulating complex fluid flow scenarios. We focus on high-Reynolds number channel flow, periodic hills, and flow over a circular cylinder, which have been widely used as benchmark cases in the field. Through comprehensive comparisons with experimental data and results from other studies, we demonstrate the practical value and reliability of our approach.

In section 2 we first introduce the QR model of LES approach before providing a discussion on the symmetry-preserving discretization and standard discretization methods in OpenFOAM. 
In section 3, we optimize the model constant by simulating the plane channel flow and comparing it to dynamic models. Subsequently, we run high Reynolds number simulations and validate the computational results against detailed DNS data from various studies ( Moser, Kim and Mansour 1999\cite{moser1999direct}; Hoyas and Jimenez 2006\cite{hoyas2006scaling}; 2008\cite{hoyas2008reynolds, jimenez2008turbulent}; and Juan et al. 2001\cite{juan2001direct}, 2003\cite{juan2003spectra}, 2004\cite{juan2004scaling}, and 2013\cite{juan2013direct}).  
In section 4, we present simulation results of periodic hills and compare them with experimental\cite{rapp2010new} and LES \cite{temmerman2001large} results from other studies.  
In section 5 we discuss the simulation results of the flow over a circular cylinder in comparison to measurement data (Lourenco and Shih 1993\cite{lourenco1994characteristics}; Ong and Wallace 1996\cite{ong1996velocity}) and numerical work by Kravchenko and Moin\cite{kravchenko2000numerical}, Mittal\cite{mittal1995large} and Breuer\cite{breuer1998large} for mean flow and turbulence quantities. Finally, we finish with the paper conclusions.

\section{Numerical Schemes}
\subsection{Minimum-dissipation model}
The dynamics of large eddies in incompressible fluid flow are governed by the following momentum and continuity equations
\begin{align}
     &\partial_t v + (v\cdot \nabla) v + \nabla p - 2\nu \nabla \cdot S(v) = -\nabla \cdot \tau (v) \\
     &\nabla \cdot v = 0
     \label{eq:1}
\end{align}

where $\nu$ stands for the viscosity and $p$ is the pressure; $S(v) = (\nabla v + \nabla v ^T)/2$ is the symmetric part of the velocity gradient. The sub-grid tensor $\tau (v)$ can be expressed as 
\begin{align*}
    \tau_{ij} &= \overline{u_i u_j} - \bar u_i \bar u_j\\
    &=\frac{1}{3} \tau_{kk} \delta_{ij} + (\tau_{ij}-\frac{1}{3}\tau_{kk}\delta_{ij})\\
    &=\frac{2}{3} k_{sgs} \delta_{ij} + (\tau_{ij}-\frac{1}{3}\tau_{kk}\delta_{ij})
\end{align*}
where $k_{sgs}=\frac{1}{2}\tau_{kk}=\frac{1}{2}(\overline{u_k u_k}-\bar u_k \bar u_k)$ is the sub-grid scale kinetic energy. The sub-grid scale stress tensor $\tau_{ij}$ is split into an isotropic part $\frac{1}{3} \tau_{kk} \delta_{ij}$ and anisotropic part $\tau_{ij}-\frac{1}{3}\tau_{kk}\delta_{ij}$. The eddy-viscosity model describes the anisotropic part of the sub-grid as
\begin{equation}
    \label{eq:3}
    \tau_{ij}-\frac{1}{3}\tau_{kk}\delta_{ij} = -2\nu_e S(v)
\end{equation}
Note that the trace of $S(v)$ is zero because $\nabla \cdot v = 0$. The coefficient $\nu_e$ is called the eddy viscosity of the model. This sub-grid model is time irreversible (for $\nu_e > 0$), forward in time it provides dissipation. The classical eddy-viscosity model\cite{smagorinsky1963general} sets the eddy viscosity equal to 
\begin{equation}
    \nu_e = C_S^2 \Delta ^2 \sqrt{4q}
\end{equation}
where $q(v)=\frac{1}{2}tr(S^2(v))$ is the second invariant of the strain-rate tensor $S(v)$. It may be remarked that $|S(v)|=\sqrt{2tr(S(v)^2)}=\sqrt{4q}$. Various values of Smagorinsky constant $C_s$ have been proposed for different cases, ranging from $C_s =0.20-0.22$ for decaying homogeneous isotropic turbulence\cite{lilly1966application}, to $C_s=0.1-0.17$ for channel flow\cite{lilly1992proposed}, and temporal mixing layers.

The first minimum-dissipation model is proposed by Verstappen\cite{verstappen2011does}. It is based on the invariants of the rate of strain tensor, and set to switch off in laminar flow and flows with negative eddy dissipation. 
Minimum-dissipation model assumes that the eddy viscosity model must keep the residual field $v'=v-\bar v$ from becoming dynamically significant. This condition is formalized by confining the sub-grid kinetic energy with Poincaré's inequality. Poincaré's inequality shows that there exists a constant $C_\Delta$, depending only on $\Omega_\Delta$, such that for every function $v$ in the Sobolev space $W^{1,2}(\Omega_\Delta)$
\begin{equation}
    \label{eq:15}
    \int_{\Omega_\Delta} \parallel v-\bar v\parallel^2 dx \leq C_\Delta \int_{\Omega_\Delta} \parallel \nabla v\parallel^2 dx
\end{equation}
where the residual field $v'=v-\bar v$ contains the eddies of size smaller than the length of the filter $\Delta$, and $\parallel \cdot \parallel = \sqrt {\langle, \rangle}$ is the standard norm of the inner product $\langle, \rangle$ on the space of real valued $L^2(\Omega_\Delta)$ functions. The Poincaré constant $C_\Delta$, independent of $v$, is equal to the inverse of the smallest non-zero eigenvalue of the dissipative operator $-\Delta = - \nabla \cdot \nabla = \nabla ^T \nabla$ on the grid cell $\Omega_\Delta$\cite{courant2008methods}. Here, it uses $-\nabla = \nabla^T$ for the $L^2(\Omega_\Delta)$ inner product and periodic domain $\Omega_\Delta$. For convex domains, the Poincaré constant is given by $C_\Delta = (\Delta/ \pi)^2$\cite{payne1960optimal}. Poincaré's inequality shows that the kinetic energy of residual field $v'$ is bounded by a constant times the velocity gradient energy
\begin{equation}
    \int_{\Omega_\Delta}\frac{1}{2}\parallel v'\parallel^2 dx \leq C_\Delta \int_{\Omega_\Delta} \frac{1}{2}\parallel \nabla v \parallel ^2 dx
\end{equation}

The evolution of velocity gradient energy can be expressed by taking the $L^2$ inner product with $\nabla^2 v$. Integration by part gives
\begin{align}
\label{eq:16}
    \frac{d}{dt} \int_{\Omega_\Delta} \frac{1}{2} \parallel \nabla v \parallel ^2 dx = -\nu \int_{\Omega_\Delta} \parallel \nabla^2 v\parallel ^2 dx + \int_{\Omega_\Delta} (v \cdot \nabla) v \cdot \Delta v dx -\nu_e \int_{\Omega_\Delta} \parallel \nabla^2 v \parallel^2 dx
\end{align}
where the boundary terms that result from the integration by parts vanish because $\Omega_\Delta$  is a periodic box. The second term in the right-hand side of the equation (\ref{eq:16}) represents the creation of velocity gradient energy by the convective term in Navier-Stokes equations. It can be expressed in the form of $r(v)=-tr(S^3(v))/3= -det S(v)$, the third invariant of strain-rate tensor $S(v)$. We suppose that the eddy viscosity and molecular viscosity are constant over a grid cell. The third term in equation (\ref{eq:16}) is the dissipation caused by eddy viscosity which can be expressed in the form of $q(\omega)=tr(S^2(\omega))/2$, the non-zero second invariant of strain-rate tensor $S(\omega)$, where $\omega$ denotes the vorticity, $\omega =\nabla \times v$.  Please refer to \cite{verstappen2011does,verstappen2018much,rozema2015minimum} for the details.

Introducing $r(v)$ and $q(v)$ into the evolution of velocity gradient energy (\ref{eq:16}), we obtain
\begin{align}
    \label{eq:25}
    \frac{d}{dt} \int_{\Omega_\Delta} \frac{1}{2} \parallel \nabla v \parallel ^2 dx 
    = -\nu \int_{\Omega_\Delta} \parallel \nabla^2 v\parallel ^2dx + 4 \int_{\Omega_\Delta} r(v) dx -4\nu_e \int_{\Omega_\Delta} q(\omega) dx
\end{align}
Now suppose the eddy viscosity is taken such that the last two terms in the RHS of equation (\ref{eq:25}) cancel each other out
\begin{equation}
    \label{eq:26}
    \int_{\Omega_\Delta} r(v) dx = \nu_e \int_{\Omega_\Delta} q(\omega) dx
\end{equation}
Then we obtain 
\begin{equation}
    \label{eq:27}
    \frac{d}{dt} \int_{\Omega_\Delta} \frac{1}{2} \parallel \nabla v \parallel ^2 dx = -\nu \int_{\Omega_\Delta} \parallel \nabla^2 v\parallel ^2dx
\end{equation}
Applying Poincaré's inequality and Gronwall's lemma  to the right-hand side of the above gives
\begin{align}
    \int_{\Omega_\Delta} \parallel v' \parallel ^2 (x,t) dx \leq C_\Delta \int_{\Omega_\Delta} \parallel \nabla v \parallel ^2 (x,t) dx \leq C_\Delta e^{\frac{-2 \nu t}{C_\Delta}} \int_{\Omega_\Delta} \parallel \nabla v'\parallel ^2 (x,0) dx
\end{align}
The energy of the sub-grid scale decays at least as fast as $C_\Delta e^{\frac{-2 \nu t}{C_\Delta}}$, for any filter length $\Delta$. So we can keep the sub-filter component $v'$ under control with the help of equation (\ref{eq:26}). The minimum eddy dissipation needs to satisfy the dissipation condition (\ref{eq:27}).

The right-hand side, $q(\omega)$ in Equation (\ref{eq:26})  can be expressed in $q(v)$:
\begin{align}
    \label{eq:29}
    \int_{\Omega_\Delta} q(\omega) dx = \frac{1}{4}\int_{\Omega_\Delta} |\nabla \omega|^2 dx \nonumber=\frac{1}{4}\int_{\Omega_\Delta} \omega \cdot (-\Delta)\omega dx \nonumber = \frac{\int_{\Omega_\Delta} \omega \cdot (-\Delta)\omega dx}{\int_{\Omega_\Delta} \omega \cdot \omega dx} \cdot \int_{\Omega_\Delta} q(v) dx 
\end{align}

Thus Eq.(\ref{eq:26}) becomes
\begin{equation}
    \nu_e = \frac{\int_{\Omega_\Delta} \omega \cdot \omega dx}{\int_{\Omega_\Delta} \omega \cdot -\Delta \omega dx} \cdot \frac{\int_{\Omega_\Delta} r(v) dx}{\int_{\Omega_\Delta}q(v) dx}
\end{equation}
The first fraction in the above right-hand side is at most $C_\Delta$, i.e. one over the smallest eigenvalue of $-\Delta$ on $\Omega_\Delta$.

Thus we take $ \nu_e \int_{\Omega_\Delta} q(v) dx = C_\Delta \int_{\Omega_\Delta} r(v) dx $. 
This equality ensures that the sub-grid scales are dynamically insignificantly, meaning that their energy is bounded by equation (\ref{eq:27}) where the energy of sub-grid scales $\int_{\Omega_\Delta} \parallel v'\parallel ^2 (x,t) d x$ decays at least as fast as the $C_\Delta e^{\frac{-2\nu t}{C_\Delta}}$, for any filter length $\Delta$. Hence, the minimum amount of eddy viscosity needed to ensure that the nonlinear production is counteracted is given by 
\begin{equation}
    \label{eq:30}
    \nu_e = C_\Delta \frac{\overline{|r(v)|}}{\overline {q(v)}} 
\end{equation}
where the absolute value of $r(v)$ is taken to make sure that the eddy viscosity is non-negative, $\overline{q(v)}$ and $\overline{r(v)}$ are grid cell average of second and third invariant of the rate of the strain tensor, respectively. In practice, the grid cell average of invariants is approximated by mid-point integration. This gives the QR model 
\begin{equation}
    \label{eq:31}
    \tau -\frac{1}{3} tr(\tau) I = -2\nu_e S(v) = -2 C_\Delta \frac{|r(v)|}{q(v)} S(v)
\end{equation}

\subsubsection{Comparison of Reynolds stress}

Since the QR model is traceless, only the deviatoric Reynolds stresses can be reconstructed and directly compared with DNS and experimental data\cite{winckelmans2002comparison}. The comparison is carried out via 
\begin{equation}
    R_{ij}^{DNS,dev} \approx   R_{ij}^{LES,dev} + \langle \tau_{ij}^{SGS,dev} \rangle,
\end{equation}
where the $\langle \tau_{ij}^{SGS,dev} \rangle$ is the averaged deviatoric SGS tensor and $R_{ij}^{dev} $ is the deviatoric Reynolds stress tensor. Here, the Reynolds stress tensor is defined as 
\begin{equation*}
    R_{ij}=\langle u_i u_j \rangle - \langle u_i \rangle \langle u_i \rangle = \langle u_i'u_j' \rangle,
\end{equation*}
where $u_i$ represents the velocity vector in DNS simulation and the coarse grid velocity vector in LES simulation.
Another way to take the contribution of the sub-grid scale into account is by reconstructing the turbulent kinetic energy from the modified pressure.
\begin{align}
    R_{ij}^{DNS} &\approx   R_{ij}^{LES} + (\langle \tau_{ij}^{SGS,dev} \rangle + \frac{2}{3} \langle \bar k_{sgs} \rangle \delta_{ij})\\
    &\approx  R_{ij}^{LES} + (-2\nu_{e} S(v) + \frac{2}{3} \langle \bar k_{sgs} \rangle \delta_{ij}),
\end{align}
since the trace of the SGS stress tensor (sometimes referred to as the sub-grid kinetic energy) has been lumped into the modified pressure.

\subsection{Numerical schemes in OpenFOAM}
Numerical simulations were performed using OpenFOAM which is based on finite volume methods for discretizing partial differential equations. The solver we use is pimpleFOAM if no other specification is provided. This solver combines the PISO (Pressure Implicit with Splitting of Operators) and SIMPLE (Semi-Implicit Method for Pressure Linked Equations) algorithms to put together the continuity equation and momentum equations.

\subsubsection{Spaital discretization}
\begin{figure}[h!]
    \centering
    \includegraphics[width=0.5\linewidth]{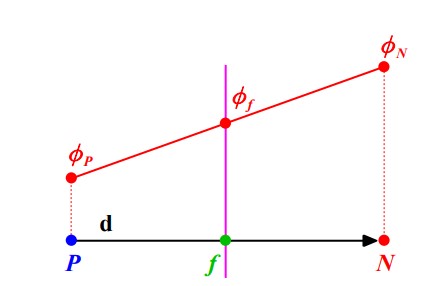}
    \captionsetup{font={footnotesize}}
    \caption{The interpolation in OpenFOAM. Here P and N denote the center of two neighbors control volumes, and f denotes the location of the interface.}
    \label{fig:discretization1}
\end{figure}

The discretization schemes are generally second-order accurate. OpenFoam adapts the collated arrangement. In this arrangement, the value of all variables is computed and stored in the center $x_P$ of the control volume $V_P$. These values are represented by a piecewise constant profile (the mean value),
\begin{equation}
    \phi(x_P)=\phi_P \approx \overline {\phi} = \frac{1}{V_P} \int_{V_P}\phi(x)dV,
    \label{eq:discretization1}
\end{equation}
where $\phi$ refers to a quantity that is discretized. 
By using Gauss or Divergence theorem, the volume integrals appearing in the governing equations are converted into surface integrals. Then, the problem reduces to interpolating cell-centered values to the face-centered values. The face values appearing in the convective and diffusive fluxes have to be computed by some form of interpolation from the centroid of the control volumes to its faces. 
The interpolation scheme applied is Gauss linear interpolation, which yields a central difference scheme on a uniform mesh, see Eq.(\ref{eq:discretization2}) and Fig.\ref{fig:discretization1}. 
\begin{equation}
    \phi_f =\alpha \phi_P +(1-\alpha)\phi_N,
    \label{eq:discretization2}
\end{equation}
where $\alpha = \frac{f-N}{P-N}$.

The spatial mesh is generated in OpenFOAM by blockMesh. The resulting mesh is 3D structured (collocated).

\subsubsection{Temporal discretization}
The first-order time derivative $\partial/\partial{t}$ is discretized with an implicit backward scheme denoted by (Eq.\ref{eq:discretization3}) if no other specification is provided. 
\begin{equation}
    \frac{3\phi^{n+1}-4\phi^{n}+\phi^{n-1}}{2\Delta t} \approx \frac{\partial}{\partial t}(\phi), 
    \label{eq:discretization3}
\end{equation}
where the $n+1$ is the value at the next time level $t + \Delta t$, $n$ is the value at the current time level, and $n-1$ is the value at the previous time level $t - \Delta t$.

\subsection{Symmetry preserving discretization scheme}
\label{sec:symmetry}
\subsubsection{Navier-Stokes equation}
The incompressible Navier-Stokes equation is written
\begin{align}
     \partial_t \vb{u} + (\vb{u}\cdot \nabla) \vb{u} - \frac{1}{Re}\nabla \cdot \nabla \vb{u} + \nabla p  = 0 ,\quad \quad \nabla \cdot \vb{u} = 0
     \label{eq:sym1}
\end{align}
where the parameter $Re$ denotes the Reynolds number.

In the absence of external sources (such as body or boundary forces), the rate of change of the total energy is neither influenced by the pressure difference nor by the convective transport; it is solely determined by dissipation. This basic physical property can be readily deduced from the symmetric properties of the differential operators in the Navier-Stokes equations (\ref{eq:sym1}).

The total energy of the flow $\vb{(u,u)}$ is defined in terms of the usual scalar product. The temporal evolution can be obtained by differentiating $\vb{(u,u)}$ with respect to time and rewriting $\partial_t\vb{u}$ with the help of Eq.(\ref{eq:sym1}). In this way, we get
\begin{equation}
    \frac{d}{dt}(\vb{u,u})= -((\vb{u}\cdot \nabla)\vb{u},\vb{u}) - (\vb{u}, (\vb{u}\cdot \nabla )\vb{u}) + \frac{1}{Re}((\nabla \cdot \nabla \vb{u},\vb{u})+ (\vb{u}, \nabla \cdot \nabla \vb{u})) -(\nabla p ,\vb{u}) - (\vb{u}, \nabla p)
\end{equation}
Integrating the linear and trilinear forms on the right-hand side by parts, ignoring any boundary contributions, we obtain 
\begin{equation}
    (\vb{u}, \nabla )^* = -(\vb{u}, \nabla) \quad \text{and} \quad \nabla ^* = -\nabla
    \label{eq:sym2}
\end{equation}

Due to these (skew-)symmetries, the convective- and pressure-dependent terms cancel and the rate of change of the total energy reduces to
\begin{equation}
    \frac{d}{dt}(\vb{u},\vb{u} )= -\frac{2}{Re}(\nabla \vb{u},\nabla \vb{u}) \leq 0
    \label{eq:sym3}
\end{equation}
In the discrete setting, the energy also evolves according to Eq.(\ref{eq:sym3}) with $\vb{u}$ replaced by the discrete velocity, and $\nabla$ by its discrete approximation, provided the discretization of the differential operator also possesses the (skew-)symmetries expressed in Eq.(\ref{eq:sym2}). Under this condition, the energy of any discrete solution remains conserved in the absence of viscosity, and it decreases over time when dissipation is present. In other words, a symmetry-preserving, spatial discretization of the Navier-Stokes equation is unconditionally stable and conservative.

\subsubsection{First-order symmetry-preserving discetization}
Consider the discretization of a first-order derivative in one spatial dimension. The Lagrangian interpolation violates the skew-symmetry of the convective operator on the nonuniform grids, and quantities conserved in the continuous formulation, like the kinetic energy, are not conserved in the discrete formulation. This leads to the fact that the energy is either systematically damped (as in the upwind methods: the convective term introduces the artificial dissipation that damps the kinetic energy) or needs to be damped explicitly to ensure stability. Nevertheless, as artificial dissipation inevitably interferes with the subtle balance between convective transport and physical dissipation, especially at the smallest scales of motion, the essence of turbulence is strained. Thus, symmetry-preserving discretization is applied. 

Consider the first-order momentum and continuous equation 
\begin{equation}
    \partial_t u +\Bar{u}\partial_xu -\partial_{xx}u/Re + \partial_x p=0,  \quad \quad \partial_x u=0
    \label{eq:sym4}
\end{equation}
where the convective transport velocity $\Bar{u}$ is taken constant, for simplicity. In matrix-vector notation the spatial discretization of Eq.(\ref{eq:sym4}) may be  written as
\begin{equation}
    \Omega_0\frac{d\vb{u_h}}{dt} +\vb{C_0}(\vb{\Bar{u}})\vb{u}_h + \vb{D}_0\vb{u}_h + \Omega_0 \vb{G}_0\vb{p}_h= 0, \quad \quad \vb{M}_0\vb{u}_h=0
    \label{eq:sym5}
\end{equation}
where the diagonal matrix $\Omega_0$ is built of the spacing of mesh: $(\Omega_0)_{i,i}=\frac{1}{2}(x_{i+1}-x_{i-1})$, the discrete velocities $u_i$ constitute the vector $\vb{u_h}$; the tridiagonal matrices $\vb{C_0}(\Bar{u})$ and $\vb{D_0}$ represent the convective and diffusive operate, respectively. 

The mass flux $\vb{\Bar{u}}$ needs to be expressed in terms of the discrete velocity $\vb{u}_h$ to close the system of Eq.\eqref{eq:sym5}. The coefficient matrix $\vb{C}_0(\vb{\Bar{u}})$ becomes a function of $\vb{u}_h$ then. Relating the mass flux $\Bar{\vb{u}}$ to the discrete velocity $\vb{u}_h$ by means of the mid-point rule gives the discrete continuity constraint, which confines the discrete velocity to $\vb{M}_0\vb{u}_h=0$, where the right-hand-side is zero only applying to the impervious or periodical boundaries.

The pressure gradient is discretized with the help of the symmetry relation \eqref{eq:sym2}. According to Eq.\eqref{eq:sym2} the continuous gradient operator is equal to the negative of the transpose of the divergence, i.e., any velocity and pressure fields satisfy $(\nabla p,\vb{u})=-(p,\nabla \cdot \vb{u})$. This relation also holds for the discrete pressure $\vb{p}_h$ and the discrete pressure gradient $\vb{G}_0 \vb{p}_h$,  that is 
 \begin{equation}
     (\vb{G}_0\vb{p}_h)^*\Omega_0 \vb{u}_h =\vb{p}_h^* \vb{G}_0^* \Omega_0 \vb{u}_h=-\vb{p}_h^*\vb{M}_0\vb{u}_h,
 \end{equation}
 if the gradient operator is approximated by
 \begin{equation}
        \vb{G}_0= -\Omega^{-1}_0 \vb{M}_0^*.
        \label{eq:sym10}
 \end{equation}
Note that the gradient matrix, describing the integration of the pressure over the control volumes $\Omega$, is given by $-\vb{M}_0^*$. Because the discrete pressure gradient inherits also the boundary condition from the discrete divergence, we need not specify boundary conditions for the pressure.

In the absence of diffusion, that is for $\vb{D_0}=0$, the energy $\norm{\vb{u_h}}^2=\vb{u_h^*}\Omega_0\vb{u_h}$ of any solution $\vb{u_h}$ of the dynamic system of \eqref{eq:sym5} is conserved if and only if the right-hand side of 
\begin{equation}
    \frac{d}{dt}\norm{\vb{u_h}}^2 = -\vb{u_h^*}(\vb{C_0}(\Bar{u})+\vb{C_0^*}(\Bar{u}))\vb{u_h} +\vb{u}_h^*(\vb{M}_0^*\vb{p}_h) + (\vb{M}_0^* \vb{p}_h)^* \vb{u}_h\nonumber
\end{equation}
is zero. This property holds (for any $\vb{u}_h$) if and only if the coefficient matrix $\vb{C}_0(\Bar{u})$ is skew-symmetric,
\begin{equation}
    \vb{C}_0(\Bar{u}) + \vb{C}_0^*(\Bar{u})=0,
    \label{eq:sym6}
\end{equation}
i.e., the discrete operator has to inherit the skew-symmetry of the continuous convective derivative $(\vb{u}\cdot\nabla)$.

The skew-symmetry condition \eqref{eq:sym6} can be satisfied if the interpolation weights of the adjacent discrete variables are taken equal to $1/2$, hence the symmetry-preserving discretization gives
\begin{equation}
    \Bar{u}\partial_x u(x_i) \approx \Bar{u} \frac{u_{i+1}-u_{i-1}}{x_{i+1}-x_{i-1}} = (\Omega_0^{-1} \vb{C_0}(\Bar{u}) \vb{u_h})_i
    \label{eq:sym7}
\end{equation}
The entries of the tridiagonal matrix $\vb{C_0}(\Bar{u})$ are given by $\vb{C_0}(\Bar{u})_{i,i-1}=-\frac{1}{2}\Bar{u}$, $\vb{C_0}(\Bar{u})_{i,i}=0$ and $\vb{C_0}(\Bar{u})_{i,i+1}=\frac{1}{2}\Bar{u}$. Manteufel and White\cite{manteufelnumerical1986} have rigorously proven that the approximation (\ref{eq:sym7}) yields second-order accurate solutions on uniform as well as on nonuniform meshes.

Diffusion is discretized in the same vein. The resulting coefficient matrix $\vb{D}_0$ is positive-definite, like the underlying differential operator $-\partial_{xx}$
\begin{equation}
    \vb{D}_0 = \frac{1}{Re}\Delta^*_0 \Lambda^{-1}_0 \Delta_0  \nonumber,
\end{equation}
where the difference matrix $\Lambda_0$ is defined by $(\Delta_0 \vb{u}_h)_i = u_i - u_{i-1}$, and the nonzero entries of the diagonal matrix $\Lambda_0$ reads $(\Lambda_0)_{i,i}=\delta x_i$. Now, the symmetric part of $\vb{C_0}(\Bar{u}+\vb{D_0})$ is only determined by diffusion and hence is positive-definite. The energy of any solution $\Bar{\vb{u}}_h$ of the semi-discrete system \eqref{eq:sym5} evolves like in the continuous case; compare Eq. \eqref{eq:sym3} to
\begin{equation}
    \frac{d}{dt}(\vb{u}_h^* \Omega_0\vb{u}_h) \stackrel{\eqref{eq:sym5} + \eqref{eq:sym6}}{=} -\vb{u}_h^*(\vb{D}_0+\vb{D }_0^*)\vb{u}_h\leq 0, \nonumber
\end{equation}
where the right-hand is zero if and only if $\vb{u}_h$ lies in the null space of $\vb{D}_0 + \vb{D}_0^*$. So, in conclusion, since the energy $\norm{ \vb {u}_h}^2 = (\vb{u}_h^*\Omega_0 \vb{u}_h)$ does not increase in time, a stable solution can be obtained on any grid. The higher-order symmetry-preserving discretization can be obtained in a similar way, (see the derivation by Verstappen and Veldmen\cite{verstappen2003symmetry}.  Taking all ingredients together yields the symmetry-preserving discretization of the Navier-Stokes equation in the next section.

\subsubsection{Symmetry-preserving discretization of  Navier-Stokes equation}
The semi-discrete representation of the incompressible Navier-Stokes equations is written
\begin{equation}
    \Omega\frac{d\vb{u}_h}{dt} +\vb{C}(\vb{u}_h)\vb{u}_h + \vb{D}\vb{u}_h -\vb{M}^*\vb{p}_h =0, \quad \quad \vb{M} \vb{u}_h = 0
    \label{eq:sym8}
\end{equation}
Global conservation laws invoke integrals over the flow domain. These integrals become scalar products when the flow is discretized.  For instance, the change of the total mass of the flow is discretized as a scalar product of constant vector $\vb{1}$ (where the dimension equals the number of grid cells) and the discrete mass flux $\vb{M} \vb{u}_h $. Since this scalar product is zero ($\vb{M} \vb{u}_h=0$) the total mass is conserved. 

The total amount of momentum is obtained by taking the scalar product of the velocity vector $\vb{u_h}$ with the vector $\Omega\vb{1}$ (where the constant vector now has as many entries as there are control volumes for the discrete velocity components $u_{i,j}$ and $v_{i,j}$). The evolution of the total amount of momentum follows straightforwardly from Eq.\eqref{eq:sym8}:
\begin{equation}
    \frac{d}{dt}(\vb{1}^*\Omega \vb{u}_h) = -\vb{1}^*(\vb{C}(\vb{u}_h)+\vb{D})\vb{u}_h +\vb{1^*M^*}\vb{p}_h =0.
    \label{eq:sym9}
\end{equation}
Hence, the momentum is conserved provided $\vb{(C(u}_h)+\vb{D})^*1=0$, and the law of the conservation of mass is consistently discretized, that is $\vb{M 1=0}$. The former condition may be split into two conditions, one for the convective discretization $\vb{C^* (u}_h) \vb{1}=0$, and one for the diffusive discretization $\vb{D^* 1=0}$. Moreover, we can leave the  $^* $'s away,  $\vb{C} (\vb{u}_h) \vb{1}=0$ and $\vb{D 1=0}$, since the convective matrix $\vb{C(u}_h)$ is skew-symmetry and the diffusive matrix $\vb{D}$ is symmetric. So it suffices to verify that the constant vector lies in the null space of the approximate, convective and diffusive operators.

The discretization is set up such that the evolution of the (kinetic) energy $\vb{u}_h^* \Omega \vb{u}_h$ of any solution of Eq.\eqref{eq:sym8} is governed by
\begin{equation}
    \frac{d}{dt}(\vb{u}_h^* \Omega\vb{u}_h) = -\vb{u}_h^*(\vb{C u}_h+\vb{C^* u}_h)\vb{u}_h -\vb{u}_h^*(\vb{D}+\vb{D^* })\vb{u}_h + \vb{u}_h^*\vb{(M^*p)}_h +\vb{(M}^*\vb{p}_h)^*\vb{u}_h, \nonumber
\end{equation}
where the right-hand side is negative for all $\vb{u}_h$'s, except those in the null space of $\vb{D + D^* }$. The convective cancels because the $\vb{C(u)}_h$ is skew-symmetric; the pressure terms cancel on the staggered grids (hence, cannot unstabilize the spatial discretization) because the discrete pressure gradient is related to the transpose of $\vb{M}$, see Eq.\eqref{eq:sym10}. 

So, in conclusion, for inviscid flow, the energy is conserved, whereas for viscous flow the energy $\norm{\vb{u_h}}^2=\vb{u_h^*}\Omega\vb{u_h}$ does not increase in time. This implies the symmetry-preserving discretization \eqref{eq:sym8} is stable  and conserves mass, momentum, and energy. 

\subsubsection{Solving the pressure-velocity coupling on the collocated grid. Checkerboard problem}
However, on a collocated grid, the actual velocity $\vb{u_c}$ is stored in the cell center. The velocity and pressure coupling term introduces an additional error term proportional to the third-order derivative of pressure $\tilde {\vb{p}}_c'$ to the momentum equation. This phenomenon is generally known as a checkerboard problem for pressure\cite{shashank2010co}. Trias et.al\cite{TRIAS2014246} proposed eliminating the checkerboard spurious mode without introducing any non-physical dissipation. The idea behind this approach is to use a linear shift operator to transform a cell-centered velocity $\vb{u_c}$ into a staggered one $\mathbf{u_s}$ and use a fully-conservative regularization of the convective term to restrain the production of the small motion.

\paragraph{Shift operators}
The linear shift operator is needed to relate the cell-centered velocity field to the staggered ones and vice versa. Here, the subscript $s$ denotes the variables staggered on the faces and $c$ denotes the variables cell-centered on the collated mesh. The cell-to-face linear shift operator is given by $\Gamma_{c\rightarrow s} \in\mathbb{R}^{m\times 3n}$, transforms a cell-centered velocity field into a staggered one 
\begin{equation}
    \vb{u}_s=\Gamma_{c\rightarrow s}\vb{u}_c,
\end{equation}
whereas the cell-centered fields are related to the staggered ones via the linear shift transformation $\Gamma_{s\rightarrow c} \in\mathbb{R}^{3n\times m}$,
\begin{equation}
    \vb{u}_c= \Gamma_{s\rightarrow c} \vb{u}_s.
\end{equation}
Note the general $\Gamma_{s\rightarrow c}\Gamma_{c\rightarrow s}=I$ holds only approximately, i.e. $\vb{u}_c\approx \Gamma_{s\rightarrow c} \Gamma_{c\rightarrow s}\vb{u}_c$.
The face-to-cell shift operator $\Gamma_{s\rightarrow c}$ is restricted by Eq.\eqref{eq:sym10} to guarantee the contribution of the pressure gradient term to the global kinetic energy vanishes. It can be expressed as follows 
\begin{equation}
    \Gamma_{s\rightarrow c} = (I_3 \otimes\Omega_c)^{-1}\Gamma_{c\rightarrow s}^*\Omega_s,
\end{equation}
where $I_3 \in \mathbb{R}^{3\times 3}$ is the identity matrix.

The linear shift operator is given by
\begin{equation}
    \Gamma_{c\rightarrow s}= N_s\Pi,
    \label{eq:sym11}
\end{equation}
where matrices $N_s \in \mathbb{R}^{m \times 3m}$ and $\Pi \in \mathbb{R}^{3m\times 3n}$  are given by 
\begin{equation}
    N_s = (N_{s,1}, N_{s,2}, N_{s,3}) \quad \text{and} \quad \Pi = I_3 \otimes \Pi _{c\rightarrow s}, 
    \label{eq:sym12}
\end{equation}
where $N_{s, i} \in \mathbb{R}^{m\times m}$ are diagonal matrices containing the $x_i$-spatial components of the face normal vectors,  and $\Pi _{c\rightarrow s} \in \mathbb{R}^{m\times n} $ is the operator that interpolates a cell-centered scalar field to the faces.

\paragraph{Correction of the cell-center predictor velocity}
To solve the velocity-pressure coupling, a classical fractional step projection method\cite{chorin1968numerical, yanenko1960economical,perot1993analysis} is used. For the staggered velocity field, $\vb{u}_s$, a velocity $\vb{u}_s^p$ can be uniquely decomposed into a solenoidal vector, $\vb{u}_s^{n+1}$ and a curl-free vector, expressed as the gradient of a scalar field, $\vb{G} \tilde{\vb{p}_c}'$. This decomposition is written as
\begin{equation}
    \vb{u}_s^{n+1} =\vb{u}^p_s - \vb{G}\tilde{\vb{p}_c}',
    \label{eq:sym13}
\end{equation}
taking the divergence of Eq.\eqref{eq:sym13} yields a discrete Poisson equation for $\tilde{\vb{p}_c}'$
\begin{equation}
    \vb{M}\vb{u}_s^p =\vb{M}\vb{u}_s^{n+1} + \vb{M}\vb{G}\tilde{\vb{p}_c}' \Rightarrow \vb{M}\vb{G}\tilde{\vb{p}_c}' =\vb{M}\vb{u}_s^p.
    \label{eq:sym14}
\end{equation}

The cell-center predicted velocity field, $\vb{u}^p_c$, is computed with the projection method, and then corrected to obtain the velocity at the next time-step, $\vb{u_c^{n+1}}$. Assuming $\vb{u}_c^p \approx \Gamma_{s\rightarrow c} \Gamma_{c \rightarrow s} \vb{u}_c^p$ is satisfied, the overall correction algorithm can be explicitly written by combining the expression of \eqref{eq:sym13} and the linear shift operator, i.e.,
\begin{align}
    \vb{u}_c^{n+1} &= \vb{u}_c^p+ \Gamma_{s\rightarrow c} \Omega_s^{-1}\vb{M}^*\vb{L}^{-1}\vb{M}\Gamma_{c\rightarrow s}\vb{u}_c^p \\
    &= \vb{u}_c^p +  (I_3 \otimes \Omega_c)^{-1}\Gamma_{c\rightarrow s}\vb{M}^*\vb{L}^{-1}\vb{M}\Gamma_{c\rightarrow s}\vb{u}_c^p,
\end{align}

where the $\Omega_s \in \mathbb{R}^{m\times m}$ is a diagonal matrix with staggered control volumes, $\Omega_c \in \mathbb{R}^{n\times n}$ is a diagonal matrix with cell-centered control volumes. The discrete Laplacian operator $\vb{L} \in \mathbb{R}^{n\times n}$,  given by $L \equiv -M\Omega^{-1}M^*$, is symmetric and negative-definite. The so-called checkerboard problem is related to the unrealistic component of the cell-centered velocity field that the pseudo-projection matrix cannot eliminate,
\begin{equation}
    \vb{u}_c^\ominus = \vb{u}_c^\ominus -(-(I_3\otimes\Omega_c)^{-1}\Gamma_{c\rightarrow s}\vb{M}^* L^{-1}\vb{M}\Gamma_{c\rightarrow s})\vb{u}_c^\ominus,
\end{equation}
where $\vb{u}_c^\ominus$ represents the spurious modes. These 'unrealistic' velocity components cannot be corrected unless they are explicitly removed. Trias et al.\cite{TRIAS2014246} proposed the regularization (smooth approximation) of the convective term to elucidate the origin of the 'unrealistic' velocity components while keeping the numerical solution free from the unphysical oscillations.

\subsubsection{Constructing the discrete operators on unstructured collocated mesh}

\paragraph{Skew-symmetry of the convective operator}
The skew-symmetry of the convective operator $\vb{C(u_s)}+\vb{C^*(u_s)}=0$ is verified in two steps\cite{verstappen2003symmetry}. Firstly, we consider the off-diagonal elements. The matrix $\vb{C(u_s)}$-dia($\vb{C(u_s)}$) is skew-symmetry if the interpolation weights of the adjacent discrete variables are taken equal to $1/2$, hence the discrete normal velocity $[\vb{u}_s]_f \approx \vb{u}_f \cdot \vb{n}_f$, located at the centroid of the cell faces $f$, is given by
\begin{align}
    [\vb{u}_s]_f = [\Gamma _{c\rightarrow s }\vb{u}_c]_f = \frac{1}{2}([\vb{u}_c]_{c1} +[\vb{u}_c]_{c2})\cdot \vb{n}_f, 
    \end{align}
where $c1$ and $c2$ are the cells adjacent to the face $f$. The entries of the matrix $[C(\vb{u}_s)]_k$ are equal to half of the flux through the face $f$ between neighboring cells $i$ and $j$, i.e.,
\begin{equation}
    [C(\vb{u}_s)]_k =\frac{1}{2}[\vb{u}_s]_f A_f,
\end{equation}
where $A_f$ is the area of the face $f$. Furthermore, for skew-symmetry of $\vb{u}_s$, the diagonal elements must be zero, that is,
\begin{equation}
    [C(\vb{u}_s)]_{k,k}=\frac{1}{2}\sum_{f\in F_f(i)}[\vb{u}_s]_f A_f=0,
\end{equation}
where the $F_f(i)$ is the set of faces boarding the face $f$. This condition is fulfilled because the discrete divergence of $\vb{u}_s$ vanishes. Hence, the unstructured collocated convective operator at cell $i$ is obtained as follows
\begin{equation}
    [C(\vb{u}_s)\vb{u}_c]_k = \sum_{f \in F_f(k)}\frac{1}{2}([\vb{u}_c]_{c1}+[\vb{u}_c]_{c2})[\vb{u}_s]_f A_f.
\end{equation}

\paragraph{Gradient, divergence, and Laplacian operators}
Integrating the continuity equation in \eqref{eq:sym1} over an arbitrary centered cell $k$ of volume $(\Omega_c)_{kk}$ yields
\begin{equation}
    \int_{(\Omega_c)_{k,k}} \nabla \cdot \vb{u} dV=\int_{\partial(\Omega_c)_{k,k}}\vb{u}\cdot \vb{n}dS=\sum_{f\in F_f(k)}\int_{S_f}\vb{u}\cdot \vb{n}dS,
    \label{eq:sym16}
\end{equation}
a second-order accurate discretization of Eq.\eqref{eq:sym16} is 
\begin{equation}
    \sum_{f\in F_f(k)}\int_{S_f}\vb{u}\cdot \vb{n}dS \approx \sum_{f\in F_f(k)} [\vb{u}_s]_f A_f.
\end{equation}
Therefore, the divergence operator is defined as 
\begin{equation}
    [\vb{M}\vb{u}_s]_k \equiv  \sum_{f\in F_f(k)} [\vb{u}_s]_f A_f =0.
    \label{eq:sym17}
\end{equation}

According to Eq.\eqref{eq:sym10}) the integrated pressure gradient operator, $\Omega_s \vb{G}$, equals the negative of the transpose of the divergence operator $-\vb{M}^*$. Hence, the discretization of the pressure gradient at the face $f$ follows from Eq.\eqref{eq:sym17}
\begin{equation}
    [\Omega \vb{G}\vb{p}_c]_f = (p_{c1}-p_{c2})A_f,
\end{equation}
where $c1$ and $c2$ are the cells adjacent to the face $f$. Note the discrete gradient inherits the boundary conditions from the discrete divergence, we need not specify boundary conditions for the pressure. Finally, we compute the pressure from a Poisson equation, which arises from the incompressibility constraint. The Laplacian operator is approximated by the matrix,
\begin{equation}
    \vb{L} = -\vb{M} \Omega_s^{-1} \vb{M}^*
\end{equation}
which is symmetric and negative-definite, like the continuous Laplacian operator $\Delta \equiv \nabla \cdot \nabla $.

\paragraph{Diffusive operator}
The method for discretizing  the Laplacian in the Poisson equation for the pressure is also applied to discretize the diffusive term in Navier-Stokes equations. The diffusive operator is viewed as the product of two first-order differential operators, a divergence and a gradient. The divergence is discretized and the gradient becomes the transpose of the discrete divergence 
\begin{equation}
    \vb{D}_c =-\frac{1}{Re}\vb{MG}=\frac{1}{Re}(\vb{M}\Omega_s^{-1})\vb{M}^* .
\end{equation}
This construction leads to a symmetric, positive-definite, approximation of the diffusive operator $-\nabla \cdot \nabla$. The collocated diffusive operator on a cell-centered variable $\vb{\phi}_c$ is given by,
\begin{equation}
    [\vb{D}_c\vb{\phi}_c] =\frac{1}{Re}\sum_{f\in F_f(k)}\frac{(\phi_{c1}-\phi_{c2})A_f}{\delta n_f},
\end{equation}
where the length $\delta n_f$ is an approximation of a distance between the centroid of the cell $c1$ and $c2$ given by $\delta n_f = |\vb{n}_f\cdot \overrightarrow{c1c2}|$. Then, the volume of the face-normal velocity cell at the face $f$ is defined as $(\Omega_s)_f =\delta n_f A_f$.

\subsubsection{Solver}
The main algorithm in the RKSymFoam solver consists of three nested iterative levels when implicit time discretization is applied\cite{komen2021symmetry,janneshopmanRKSymFoam}:

1. an outer loop over each RK stage $i$, indicated by $1\leq i \leq s$ (see Butcher Tableau in \ref{sec:Butcher});

2. an outer iteration loop for updating the non-linear convective term;

3. an inner PISO iteration loop for the pressure-velocity coupling.

When explicit temporal discretization is applied, one projection step is used for the pressure-velocity coupling, and no outer iterations are required for updating the convective term.

Note that the pressure solver employed in the last outer loop of the simulation should be consistent with the solver used in all the previous outer loops. In our investigation of a circular cylinder simulation at a Reynolds number of 3900, we observed notable differences in execution times. When using the GAMG solver with DICGauss-Seidel preconditioner in the previous outer loops and switching to DICPCG in the last outer loop, the execution time for one time-step was approximately 60 seconds. However, when employing GAMG with Gauss-Seidel preconditioner in all the outer loops, the execution time reduced to approximately 20 seconds.
Furthermore, Issa \cite{issa1986solution} demonstrated that the order of accuracy increases by one for each additional corrector stage. To obtain a sufficiently accurate pressure field, it is recommended to use at least two corrector stages.

\subsubsection{Temporal discretization}
The time discretization in the Eq.\eqref{eq:sym8} has to be replaced by a skew-symmetric operator to preserve the favorable conservation and stability properties for discrete-time too. This can only be achieved when the time integration is done implicitly. For the use in high Reynolds number flow simulations, the computational cost of the implicit method may be higher than the explicit method. Hence, the explicit Runge-Kutta is chosen.

However, the time step $\Delta t$ of an explicit time interpolation method for a convection-diffusion equation is typically restricted by a convective stability condition like $U\Delta t < \Delta y$ (where U denotes the absolute maximum of the velocity and $\Delta y$ stands for the spatial mesh size), and a diffusive stability condition of the form of $2\Delta t < Re \Delta y^2$. 

In practical simulations of a circular cylinder at a Reynolds number of 3900, we have found that explicit Runge-Kutta time schemes require a very small CFL (Courant-Friedrichs-Lewy) number, typically smaller than 0.0003, to ensure stability. However, the implicit backward Euler method does not impose such stringent restrictions, allowing for a time step that can be 1000 times larger, such as 0.2. Nevertheless, it should be noted that the implicit backward Euler method is only first-order accurate.

To achieve second-order accuracy while maintaining stability, we employ the Crank-Nicolson and diagonal-implicit Runge-Kutta methods in our simulations. These methods strike a balance between accuracy and stability, ensuring that the simulation remains globally second-order accurate.

\paragraph{Butcher tables specify Runge--Kutta methods}
\label{sec:Butcher}
The ordinary differential equation can be obtained by applying spatial discretization to the momentum equations \eqref{eq:sym1}
\begin{equation}
    \frac{d\vb{u}_c}{dt} = \vb{F}_c(t, \vb{u}_c)-(\nabla p)_c \quad \text{with} \quad \vb{u}_c(t^n) = \vb{u}_c^n,
    \label{eq:symT1}
\end{equation}
where $c$ indicates the cell-centered values and the pressure gradient term is not yet discretized. 
The Runge-Kutta schemes, including the explicit Rugg-Kutta (ERK) and Diagonal-implicit Runge-Kutta (DIRK), are used to discretize the temporal terms in Eq.\eqref{eq:symT1}. 
A Butcher table is a simple mnemonic device for specifying a Runge--Kutta method and has the form\cite{butcher1996history}
\begin{equation*}
  \begin{array}{c|cccc}
    c_1       & a_{1,1}     & a_{1,2}    &  \dots    & a_{1,s} \\
    c_2       & a_{2,1}     & a_{2,2}    &  \dots    & a_{2,s} \\
    \vdots    & \vdots     & \vdots     &  \ddots   & \vdots \\
    c_s       & a_{s,1}     & a_{s,2}   &   \dots    & a_{s,s} \\
    \hline
    \,        & b_1        & b_2       &  \dots     & b_s
  \end{array}
\end{equation*}
where $a_{ij}$ represents the stage weights of the stage $i$, and $c_i$ are the quadrature nodes of the schemes with  $c_i = \sum\limits_{j=1}^s a_{i,j}$, for $i=1,....s$, and $t^i=t^n+ c_i \Delta t$. Furthermore, $s$ denotes the number of stages, and $b_j$ represents the main weights of the applied Runge-Kutta scheme with $\sum\limits_{j=1}^s b_j =1$.
The explicit methods are precisely those for which the only non-zero entries in the $a_{i,j}$-part of the table lie strictly below the diagonal. Entries at or above the diagonal will cause the right-hand side of \eqref{eq:symT1} to involve $\vb{u}_c^{n+1}$, and so give an implicit method.

The intermediate solution $\tilde{\vb{u}}_c^i$ for the stage $i$ at time $t^i$ is given by 
\begin{equation}
    \tilde{\vb{u}}_c^i = \vb{u}_c^n +\Delta t \left(\sum\limits_{j=1}^i a_{ij} F_c(t^j, \tilde{\vb{u}}_c^j) - c_i (\nabla \tilde{p})_c^i \right),
\end{equation}
and the final solution $\vb{u}_c^{n+1}$ at time $t^{n+1}=t^n +\Delta t$ by
\begin{equation}
    \vb{u}_c^{n+1} = \vb{u}_c^n +\Delta t \left(\sum\limits_{j=1}^i b_j F_c(t^j, \tilde{\vb{u}}_c^j) -  (\nabla p )_c^{n+1} \right).
\end{equation}

\paragraph{The pressure-velocity coupling for explicit time integration}
The forward-Euler is explained as an example since the higher-order explicit Runge-Kutta integration schemes essentially consist of  a sequence of forward-Euler stages. An intermediate velocity field $\vb{u}_c^p$ is computed from the following predictor step
\begin{equation}
    \frac{\vb{u}_c^p-\vb{u}_c^n}{\Delta t^n} = \vb{H}(\vb{u}_s^n )\vb{u}_c^n-\vb{G} \vb{p}_c^p,
    \label{eq:sym18}
\end{equation}
where $\vb{H}(\vb{u}_s^n )\vb{u}_c^n \equiv -\Omega_c^{-1}(\vb{C}(\vb{u}_s^n)\vb{u}_c^n+\vb{D}\vb{u}_c^n)$ , with $\vb{H}(\vb{u}_s^n) \in \mathbb{R}^{3n\times 3n}$, and $\Delta t^n$ is the time between time levels $n$ and $n+1$. Since the intermediate velocity is not divergence-free, the final values for the time-step $n+1$ are obtained by adding the following corrections to the intermediate values: $\vb{u}_c^{n+1} = \vb{u}_c^p +\vb{u}_c'$ and $\vb{p}_c^{n+1}=\vb{p}_c^p + \tilde{\vb{p}_c}'$. The following relation holds between the velocity and pressure correction $\vb{u}_c'$ and $\tilde{\vb{p}_c}'$
\begin{equation}
    \vb{u}_c' = -\Delta t \vb{G}\tilde{\vb{p}_c}'.
    \label{eq:sym19}
\end{equation}
Once the pressure correction $\tilde{\vb{p}}'$ is obtained from the Possion equation \eqref{eq:sym14}, the velocity correction $\vb{u}'$ can be calculated using Eq.\eqref{eq:sym19}. Finally, the new velocity and pressure fields at the next time step can be calculated using the velocity and pressure corrections $\vb{u}_c'$ and $\tilde{\vb{p}_c}'$.

\paragraph{Pressure-velocity coupling for implicit time integration}
For the implicit time integration, the PISO (Pressure-Implicit with Splitting of Operator) approach is used as a base of the RKSymFoam solver. The backward Euler time integration is explained as an example.  The PISO method consists of one predictor step followed by $n_{corr}$ corrector steps (or inner iterations) shown as follow\cite{issa1986solution}:

1)Predictor step: The first intermediate velocity $\vb{u}_c^1$ is computed from the following predictor equation
\begin{equation}
    \frac{\vb{u}_c^1-\vb{u}_c^n}{\Delta t^n} = \vb{H}(\vb{u}_s^p )\vb{u}_c^p-\vb{G} \vb{p}_c^p,
    \label{eq:sym20}
\end{equation}
where $\vb{H}(\vb{u}_s^p )\vb{u}_c^p = \vb{H}(\vb{u}_s^n )\vb{u}_c^n $ is applied in order to avoid implicit treatment of the convective and diffusive terms. The obtained first intermediate velocity will generally not be divergence-free. Hence, the corrector steps are performed subsequently.

2) corrector steps: In each correct step, a new pressure field $\vb{p}_c^k$ and a corresponding revised velocity $\vb{u}_c^{k+1}$ which is divergence-free $\vb{M}\Gamma _{c \rightarrow s}\vb{u}_c^{k+1} =\vb{0}_{c,n}$  are determined. To improve the stability of the momentum equation in the corrector step, the operator is split into a diagonal part $\vb{A}^d(\vb{u}_s^n)$ which operates on $\vb{u}_c^{k+1}$ and an off-diagonal parts $\vb{H}^{od}(\vb{u}_s^n)$ which operates on the $\vb{u}_c^{k}$, yielding 
\begin{equation}
    \vb{u}_c^{k+1} =\vb{B}^{-1}(\Delta t \vb{H}^{od}(\vb{u}_s^n)\vb{u}_c^k+\vb{u}_c^n) -\Delta t \vb{B}^{-1}\vb{G}\vb{p}_c^k,\quad \textrm{with} \quad
    k=1,2,...,n_{corr}, \label{eq:symT2}
\end{equation}
where $\vb{B}=\vb{I}-\Delta t^n A^d(\vb{u}_s^n) \in \mathbb{R}^{3n\times 3n}$. A Poisson equation for pressure $\vb{p}_c^k $ can be obtained by taking the divergence of the Eq.\eqref{eq:symT2} and by using the $\vb{M}\Gamma _{c\rightarrow s}\vb{u}_c^{k+1}=\vb{0}_{c,n}$
\begin{equation}
    \vb{M}\vb{B}_S^{-1}\Omega_s^{-1}\vb{M}^*\vb{p}_c^k= \frac{1}{\Delta t}\vb{M}\Gamma_{c \rightarrow s}\vb{B}^{-1}(\Delta t \vb{H}^{od}(\vb{u}_s^n)\vb{u}_c^k +\vb{u}_c^n), \quad \text{with} \quad k=1,2,...,n_{corr},
    \label{eq:symT3}
\end{equation}
where $\Gamma_{c \rightarrow s}\vb{B}^{-1}\Gamma_{s \rightarrow c} \approx \vb{B}_s^{-1}$, where $\vb{B}_s$ is a square diagonal matrix. The Laplacian term for the implicit scheme is $\vb{L}=-\vb{M}\vb{B}_s^{-1}\Omega_s^{-1}\vb{M}^*$.
The pressure correction $\tilde{\vb{p}}_c'^{k}$ can be obtained from the following Poisson equation
\begin{equation}
    \vb{M}\vb{B}_S^{-1}\Omega_s^{-1}\vb{M}^*\tilde{\vb{p}}_c'^{k}= \frac{1}{\Delta t}\vb{M}\Gamma_{c \rightarrow s}\vb{B}^{-1}(\Delta t \vb{H}^{od}(\vb{u}_s^n)\vb{u}_c^k +\vb{u}_c^n -\vb{G}\vb{p}_c^n), \quad \text{with} \quad k=1,2,...,n_{corr},
    \label{eq:symT4}
\end{equation}

Once the pressure field $\vb{p}_c^k$ is obtained from the Poisson equation \eqref{eq:symT3}, the corresponding revised velocity  can be computed from Eq.\eqref{eq:symT2}. This process is repeated for $n_{corr}$ iterations until an inner iteration criterion is satisfied. Then the inner PISO iteration process is finalized by updating the velocity which is used to resolve the non-linearity in the convective flux term in the operator $\vb{H}$ with the new velocity $\vb{u}_c^{n_{corr}}$ for the next outer iterator.

\section{Channel flow}
Turbulent channel flow is one of the most fundamental wall-bounded shear flows and it has been widely used to study the structure of near-wall turbulence \cite{moser1999direct}. The numerical investigations of the minimum--dissipation model and dynamic minimum-dissipation models applied to channel flow are presented, for friction Reynolds numbers up to $Re_\tau = 2000$ (based on the half channel width). The model contribution on different mesh resolutions is studied and the symmetry-preserving discretization is compared with the standard OpenFOAM discretization at $Re_\tau =1000$. 
\subsection{Physical and Numerical Domain}
\begin{figure}[htbp!]
    \centering
    \includegraphics[width = 0.48\linewidth]{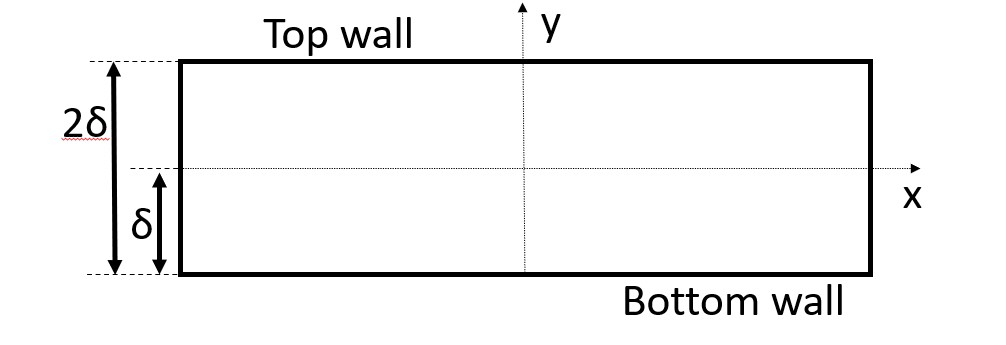}
    \includegraphics[width = 0.48\linewidth]{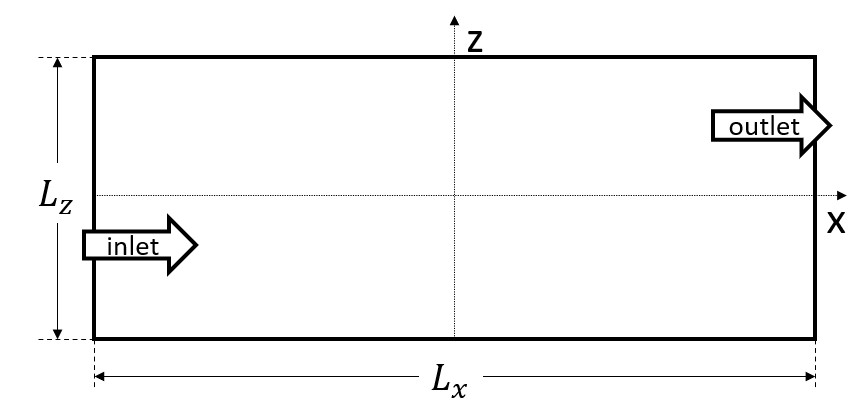}
    \caption{Numerical domain in xy-plane(left) and xz-plane (right)}
    \label{fig:channel1}
\end{figure}

The Cartesian coordinate system is shown in Figure \ref{fig:channel1}. The x-axis coincides with the direction of the mean flow  and is referred to as the streamwise direction. The y-axis is the wall-normal direction. The z-axis is orthogonal to both x- and y-axis and is called the spanwise direction. The distributions of mean velocity and Reynolds stress components in the wall-normal direction are matters of interest to researchers and engineers. Many DNS calculations have been carried out and produced a considerable amount of informative data that can be used to test the quality of LES turbulence models. 

The physical and numerical parameters for the test cases are given in Table \ref{tab:channel1}. The dimension of the channel is chosen to ensure that the turbulence fluctuations are uncorrelated at one half-period in the homogeneous directions. 
The Reynolds number based on the bulk mean velocity and the half channel width is given by $Re_b=U_b \delta/\nu$, where $\delta$ is the channel half width, $\nu$ is the fluid viscosity and $U_b$ is the mean streamwise velocity. Furthermore, the results are normalized in wall units, indicated by the plus sign (the friction Reynolds number, coordinates and the friction velocity are defined by $Re_{\tau} = \delta ^+ = \delta u_{\tau}/\nu$, $y^+=y\nu/u_{\tau}$, and $u^+=u/u_{\tau}$), respectively.

\begin{table}[htbp!]
    \centering
    \setlength{\heavyrulewidth}{1.5pt}
    \setlength{\abovetopsep}{4pt}
    \footnotesize
    \begin{tabular}{ccccccccccc}
    \toprule
Case  &  $Re_\tau$  &  $Re_b$   &  $U_b$  &  $\nu $  &  $L_x \times L_z$  & $ N_x \times N_y \times N_z$  &  $\Delta x^+$  &  $\Delta y_c^+$  &  $\Delta y_w^+$  &  $\Delta z^+$\\
\midrule
              \multicolumn{11}{c}{ \raggedleft  Simulations in this paper}\\
\midrule
QR180  &  180  &  2800    &  0.1335  &  $4.7678 \times 10^{-5}$  &  $4\delta \times 2\delta $  &  $40\times50\times30$  &  18  &  18.49  &  1.728  &  12\\
QR500  &  500  &  9159   &  0.1335  &  $1.45763 \times 10^{-5}$  &  $2\pi \times\pi$  &  $128\times128\times128$  &    &    &    & \\
QR550  &  550  &  10000  &  1.0  &  $1.000 \times 10^{-4}$  &  $8\pi \times 3\pi$  &  $256\times256\times256 $ &    &    &    &  \\
QR1006  &  1006 & 20519  & 0.1335 & $6.357 \times 10^{-6}$ & $2\pi \times \pi$ & $128 \times 128 \times 128$ & 49 & 16 &  1.95  &  24\\
QR1906  &  1906  &  42971    &  0.1335  &  $3.07052 \times 10^{-6}$ &  $2\pi \times \pi$  &  $256\times256\times256$  &   47 & 14.4 &  1.8 &  23\\
\midrule
   \multicolumn{11}{c}{ \raggedleft Reference data}   \\
\midrule
LM180  &  182  &  2857   &  1.0  &  $3.50\times 10^{-4}$ &  $8\pi \times 3\pi$  &  $1024\times192\times512$  &  4.5  &  3.4  &  0.074  &  3.1\\
LM500  &  502  &      &  1.0  &  $1.00 \times 10^{-4}$  &  $20\pi \times 5\pi $ &  $3072\times256\times1536$  &  10.3  &  6.34  & 0.04 &5.1 \\
LM550  &  543  &  10000   &  1.0  &  $1.00 \times 10^{-4}$  &  $8\pi \times 3\pi$  &  $1536\times384\times1024$  &  8.9  &  4.5  &  0.019  &  5.0\\
LM1000 &  1000.5 &  20000  &  1.0 &  $5.000\times 10^{-5}$ &  $8\pi \times 3\pi $ &  $2304\times 512\times 2048$ & & \\
HJ2000  &  2003  &  43650   &  0.89  &  $2.06186 \times 10^{-5}$  &  $8\pi \times 3\pi$  &  $4086\times 317 \times 3090$  &  12.3  &  8.9  &  0.323  &  6.1  \\
    \bottomrule
    \end{tabular}
    \captionsetup{font={footnotesize}}
    \caption{ Simulation parameters for the large-eddy simulations for the QR model and DNS reference data. Here $\Delta x^+= \Delta x/(\nu/u_\tau)$ and $\Delta z^+=\Delta z/(\nu/u_\tau)$  are the resolutions in streamwise and spanwise directions in wall units. $\Delta y_c^+$ is the maximum spacing (at the centreline of the channel) in wall units and $\Delta y_w^+$ is the $y$ resolution at the first mesh point away from the wall. $L_x$ and $L_z$ are the periodic streamwise and spanwise dimensions of the numerical domain, and $\delta $ is the channel half-width. $N_x, N_y$, and $N_z$ are the number of collocation points in each direction.}
    \label{tab:channel1}
\end{table}

 Fully developed channel flow is homogeneous in the streamwise and spanwise directions, hence periodic boundary conditions are used in these directions. The boundary conditions on the wall are no-slip for velocity, zero pressure gradient, and vanishing eddy viscosity. The mesh distribution is uniform in the streamwise and spanwise directions and stretched in the wall-normal direction (clustered near the walls). The velocity field is initialized using the minimum-dissipation model’s results obtained on a coarser grid after 10000-time steps with the help of OpenFOAM build-in function mapFields. In this way, fewer time steps are needed before starting the averaging process. The time step is chosen so that the Courant-Friedrichs-Lewy number is less than 0.8 in every simulation. Only a few hundred-time steps are required with this method to obtain a fully developed turbulent flow. The bulk velocity and kinetic viscosity are pre-set. The friction velocity $u_\tau$ is calculated with $u_{\tau}=\sqrt{\tau_{w}/\rho}$, where $\tau_w = \nu \partial{u}/\partial {y}$ is the wall shear stress.   

\subsection{Results and discussion}

The results at Reynolds numbers $Re_\tau= \frac{\delta u_\tau}{\nu}\approx 180,550,1009$ and $2000$ are compared to the DNS data from Moser et al. Scaling parameters of wall units in the results are adjusted so that the numerical results correspond to the reference data.  Mean streamwise velocity has been analyzed under five different error quantification methods, including mean square error, absolute error, maximum absolute error, slope error, and the discrepancy between DNS and QR van Karman constant,  to find the best model constant. Note that using the slope of the mean velocity profile to quantify errors is not an accurate way because too few data points (about 5 points) are located in the linear region.

\subsubsection{Optimal QR model constant at Re$_\tau$ = 180}

\begin{figure}[htbp!]
    \centering
    \includegraphics[width=0.4\linewidth]{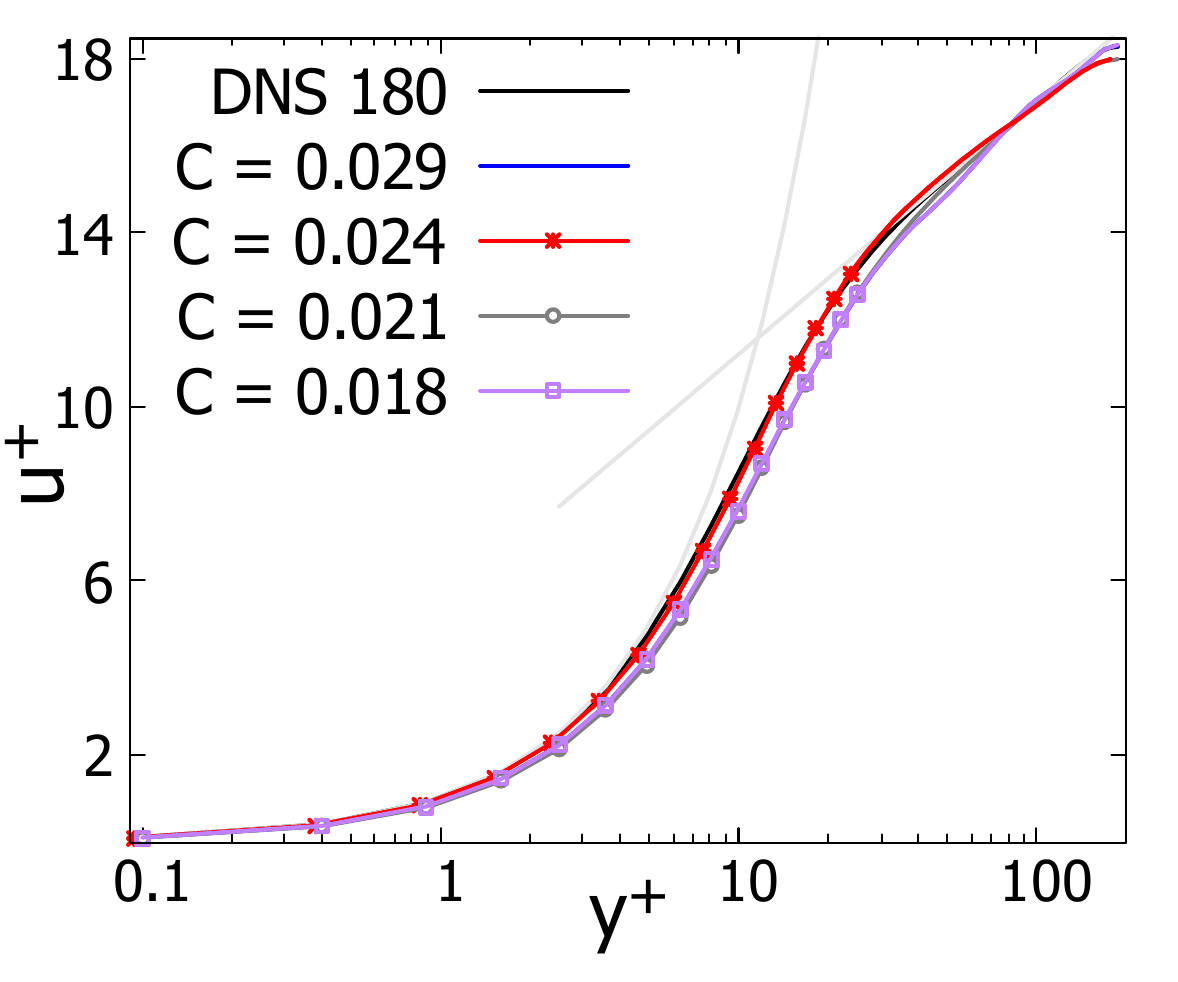}
    \captionsetup{font={footnotesize}}
    \caption{Normalized mean streamwise velocity against wall distance in wall units at $Re_\tau=180$, the computational domain is $4\delta \times 2 \times 2\delta$, the grid resolution is $40\times 50\times 30$. Four different model constants are compared with the DNS data: C = 0.018, 0.021, 0.024, and 0.029.}
    \label{fig:channel2}
\end{figure}
In this part, QR models with different model constants are applied to the channel flow at $Re_\tau=180$. Note that the model constant used in the minimum-dissipation model corresponds to the square of the Smagorinsky model constant.  The computational domain is $4\delta \times 2 \times 2\delta$, the grid resolution is $40\times 50\times 30$, the normalized uniform grid point in streamwise direction is $\Delta x^+=18$ and in spanwise direction is $\Delta z^+=12$, the simple grading expansion in wall-normal direction is around 10, the first normalized wall-normal grid point next to the wall is  $\Delta y_w^+=1.728$. The time step for simulation is $\Delta t^+=36$.

Figure  \ref{fig:channel2} shows the normalized mean streamwise velocity against the wall distance in the wall unit. As we can see from Figure \ref{fig:channel2}, a small value, for instance, C = 0.018, underestimates the mean velocity in the whole channel. This is also the case for C = 0.029. While the medium value of C = 0.024 is precisely in line with the DNS data. 

The errors quantified with the five measurements are summarised in Table \ref{tab:channel2}. What stands out in this table is that the model constant C = 0.024 gives the smallest error (mean square error, absolute error, maximum absolute error, and slope error). The value of C = 0.023, however, gives the lowest error if the van Karman constant is used to quantify the error. Note that this measurement considers only the difference in the logarithmic region ($y^+>30$).  

These findings indicate that  C = 0.023 is more accurate in the log wall region. In the near wall region, the optimal minimum-dissipation model constant is C = 0.024. According to the literature, the best value of the Smagorinsky model is between $C_s=0.1$ and $C_s=0.2$.  Thus, the optimal constant of the QR model is found in the range of $C_s^2$. 
\begin{table}[!htbp]
    \centering
    \setlength{\heavyrulewidth}{1.5pt}
    \setlength{\abovetopsep}{4pt}
    \footnotesize
    \begin{tabular}{ccccccccccc}
    \toprule
Measurement & 0.095 & 0.063 & 0.0289 & 0.027 & 0.025 & 0.024 & 0.023 & 0.022 & 0.021 & 0.01    \\
\midrule
MSE & 0.1172 & 0.1169 & 0.1050 & 0.0863 & 0.0665 & 0.0611 & 0.0737 & 0.0953 & 0.1122 & 0.0735 \\
AbsE & 0.2192 & 0.2147 & 0.2133 & 0.2715 & 0.2163 & 0.2019 & 0.2021 & 0.2564 & 0.2367 & 0.2281\\
MaxAbsE & 0.9642 & 1,0935 & 0.8798 & 0.7197 & 0.6292 & 0.5646 & 0.8049 & 0.7346 & 0.9677 & 0.6619\\
meanSE & 0.1026 & 0.1143 & 0.0980 & 0.0980 & 0.0897 & 0.0779 & 0.1103 & 0.1006 & 0.1272 & 0.0944\\
varSE & 0.0004 & 0.0006 & 0.0004 & 0.0002 & 0.0002 & 0.0002 & 0.0002 & 0.0002 & 0.0002 & 0.0002\\
stdSE & 0.0212 & 0.0241 & 0.0190 & 0.0137 & 0.0139 & 0.0130 & 0.0144 & 0.0138 & 0.0145 & 0.0139\\
meanVK & 0.3288 & 0.4890 & 0.3198 & 0.1520 & 0.1524 & 0.1612 & 0.1221 & 0.2139 & 0.1700 & 0.1648\\
varVK & - & 0.4890 & 0.3198 & 0.1520 & 0.1524 & 0.1612 & 0.1221 & 0.2139 & 0.1700 & 0.1648\\
stdVK & 0.0057 & 0.0070 & 0.0057 & 0.0039 & 0.0039 & 0.0040 & 0.0035 & 0.0046 & 0.0041 & 0.0041\\
$U_b^+$ & 15,396 & 15.376 & 15,396 & 15.318 & 15.4099 & 15.392 & 15.439 & 15.341 & 15.3531 & 15.3617\\
vk & 0.3918 & 0.4005 & 0.3932 & 0.4024 & 0.4032 & 0.4021 & 0.4053 & 0.4032 & 0.4021 & 0.4031 \\
    \bottomrule
    \end{tabular}
    \captionsetup{font={footnotesize}}
    \caption{Error quantification based on mean streamwise velocity profile at $Re_\tau= 180.3767$. MSE: mean square error; AbsE: absolute error; MaxAbsE: maximum absolute error; meanSE: mean of slope error; varSE: variance of slope error; stdSE: standard deviation of slope error;  meanVK: mean of divergence among DNS and QR van Karman constant; varVK: variance of divergence among DNS and QR van Karman constant; stdVK: standard deviation of divergence between DNS and QR van Karman constant; vK: van Karman constant extracted from QR simulations. The van Karman constant and normalized mean streamwise velocity calculated by the DNS data are 0.3959 and $U_b^+=15.5898$, respectively.}
    \label{tab:channel2}
\end{table}

\subsubsection{QR model in comparison to dynamic models at Re$_\tau$ = 180}
\begin{figure}[ht!]
    \centering
    \includegraphics[width=0.5\linewidth]{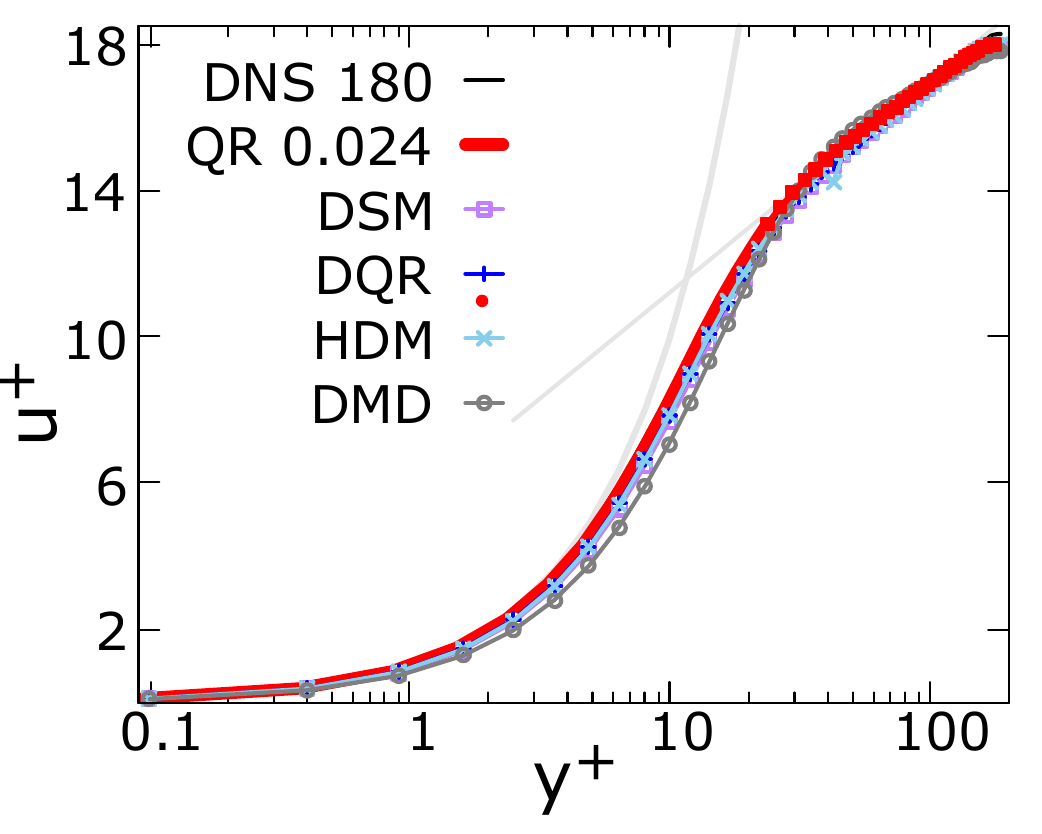}
    \captionsetup{font={footnotesize}}
    \caption{Normalized streamwise velocity $u^+$ against wall distance in wall units $y^+$. At friction Reynolds number $Re_\tau=180$, grid resolution is  $40\times 50\times 30$, size of channel is $4\delta \times 2 \times 2\delta$. Five models are compared to DNS data (black-solid line): QR 0.024: QR model with C=0.024 (red line); DQR: dynamic QR model (blue line); HDM: hybrid dynamic model (sky-blue line); DSM: dynamic Smagorinsky model (purple line), and DMD: dynamic minimum-dissipation model (grey line).}
    \label{fig:channel4}
\end{figure}

\begin{table}[htbp!]
    \centering
    \footnotesize
    \begin{tabular}{cccccc}
    \toprule
Measurement & DMD & DSM & DQR & HDM & QR 0.024 \\
\midrule
MSE & 0.2169 & 0.0953 & 0.0582 & 0.0614 & 0.0611\\
AbsE & 0.3647 & 0.2629 & 0.1849 & 0.2103 & 0.2019\\
MaxAbsE & 1,3887 & 0.7928 & 0.6070 & 0.6218 & 0.5646\\
meanSE & 0.1937 & 0.1093 & 0.0849 & 0.0910 & 0.0779\\
varSE & 0.0006 & 0.0002 & 0.0002 & 0.0002 & 0.0002\\
stdSE & 0.0239 & 0.0147 & 0.0132 & 0.0147 & 0.0130\\
meanVK & 0.0059 & 0.0043 & 0.0029 & 0.0037 & 0.0033\\
varVK & 0.3272 & 0.1576 & 0.1428 & 0.1854 & 0.1612\\
stdVK & 0.0057 & 0.0040 & 0.0038 & 0.0043 & 0.0040\\
$U_b^+ $& 15,3885 & 15,3269 & 15,4049 & 15,3796 & 15,392\\
vk & 0.4078 & 0.4041 & 0.4062 & 0.4019 & 0.4021\\
\bottomrule
    \end{tabular}
    \captionsetup{font={footnotesize}}
    \caption{Error quantification based on the mean streamwise velocity predicted by five models. The van Karman constant is 0.3959 and the normalized mean streamwise velocity is 15.5898 calculated from the DNS data. For details of the definition see Table \ref{tab:channel2}}
    \label{tab:channel4}
\end{table}
The focus of this part is on the comparison of the static QR model with C = 0.024 and four dynamic models. Figure \ref{fig:channel4} illustrates the normalized mean streamwise velocity against wall distance in wall unit at $Re_\tau=180$ with the grid resolution of $40\times 50\times 30$. And the computational domain is $4\delta \times 2 \times 2\delta$. The normalized uniform grid spacing in streamwise direction is $\Delta x^+=18$ and in spanwise direction is $\Delta z^+=12$, the simple grading expansion in wall-normal direction is around 10, the first normalized wall-normal grid height is $\Delta y_w^+=1.728$. The time step for simulation is $\Delta t^+=36$.  

As shown in Figure \ref{fig:channel4}, all LES models are very close to the DNS results in the near-wall region, especially the dynamic QR model that is overlapping with DNS data. A closer inspection of this figure reveals that in the log wall range $y^+>30$, the dynamic QR model, static QR model with constant coefficient C = 0.024, and the dynamic minimum-dissipation model are closer to DNS results. The dynamic Smagorinsky model, and hybrid dynamic model, however, are almost overlapping and apparently differ from the DNS results in the log wall region. On top of that, all LES models result in a lower center-line mean velocity compared to DNS. The reason for this is probably that the Reynolds number used here is not high enough to exhibit the logarithmic region.

Table \ref{tab:channel4} further shows that the dynamic QR model and static QR model perform more or less the same. The minor difference between the errors is insignificant to distinguish one model from another. Meanwhile, the dynamic Smagorinsky model and hybrid dynamic model perform equally less accurately, and the dynamic minimum-dissipation model has the highest error.

In conclusion, the investigation indicates that the static QR model is reliable. A properly chosen value of the QR coefficient C can provide very similar results to a dynamic model, but with a reduced computational cost. 

\subsubsection{Simulations at high Reynolds number Re$_\tau$ = 1006}
Now consider the study at high Reynolds number $Re_\tau=1006$ (based on channel half width). The results are compared to DNS data at $Re_{\tau}=1000$ from Lee and Moser et al.\cite{lee2015direct}.  
The physical and numerical parameters for the test cases are listed in Table \ref{tab:channel1}. The computational domain is $2\pi \times 2\delta \times \pi$ , the grid resolution is $128^3, 64^3, 32^3$ and $16^3$. The normalized uniform grid spacing in the streamwise direction is $\Delta x^+=49$ and in the spanwise direction is $\Delta z^+=24$. In the wall-normal direction, the mesh is non-uniform with a stretching factor of 8,  the first normalized wall-normal grid point adjacent to the wall is  $\Delta y_w^+=1.9$. The CPU time used for computing the $128^3$ case is eight hours using 32 processors on one HPC node exclusively.

\paragraph{QR Model effectivity}
\begin{table}[htbp!]
    \setlength{\heavyrulewidth}{1.5pt}
    \setlength{\abovetopsep}{4pt}
    \footnotesize
    \begin{tabular}{cccccccccc}
\toprule
Case  &  $Re_\tau$  &  $Re_b$  & $u_\tau$ &  $U_b$  &  $\nu $  &  $L_x \times L_z$  & $ N_x \times N_y \times N_z$ & Model Constant\\
\hline
16    &  1002.1  & 18375    &   0.0064    &  0.1168  & 6.357e-6	 & $2\pi \times\pi$	 & $128\times140\times128$	&    QR 0.024 \\
16NM    &  999.3  & 18376.1   &   0.0064    &  0.1168  & 6.357e-6	 & $2\pi \times\pi$	 & $128\times140\times128$	&    QR 0.000 \\
17    &  888.33  & 19185.15   &   0.00565   &  0.12196	 & 6.357e-6	 & $2\pi \times\pi$	 & $64\times64\times64$	&    QR  0.024 \\   17NM    &  912.38  & 19156.6    &   0.0058    &  0.12178 	 & 6.357e-6	 & $2\pi \times\pi$	 & $64\times64\times64$	&    QR  0.000 \\    
20    &  749.14  & 18955.72   &   0.00476   &  0.120504	 & 6.357e-6	 & $2\pi \times\pi$	 & $32\times32\times32$	&    QR  0.024 \\ 
20NM   &  787.92  & 19159.26   &   0.005008  &  0.121798  & 6.357e-6	 & $2\pi \times\pi$	 & $32\times32\times32$	&    QR  0.000 \\ 
21    &  677.18  & 18593.72   &   0.0043    &  0.118203  & 6.357e-6	 & $2\pi \times\pi$	 & $16\times16\times16$	&    QR  0.024 \\
21NM    &  757.43  & 19318.16   &   0.0048    &  0.122808	 & 6.357e-6	 & $2\pi \times\pi$	 & $16\times16\times16$	&    QR  0.000 \\
\bottomrule
\end{tabular}
\captionsetup{font={footnotesize}}
\caption{Simulation parameters for the QR model combined with standard OpenFOAM discretization using PimpleFoam solver at $Re_\tau=1000$. $L_x$ and $L_z$ are the periodic streamwise and spanwise dimensions of the numerical domain, and $\delta $ is the channel half-width. $N_x, N_y$, and $N_z$ are the number of collocation points in each direction. Model constant refers to the LES model and the model constant applied. Smag denotes the Smagorinsky model.}
\label{tab:channel5}
\end{table}

\begin{figure}
    \centering
    \includegraphics[trim = 20 0 30 25, clip, width=0.45\linewidth]{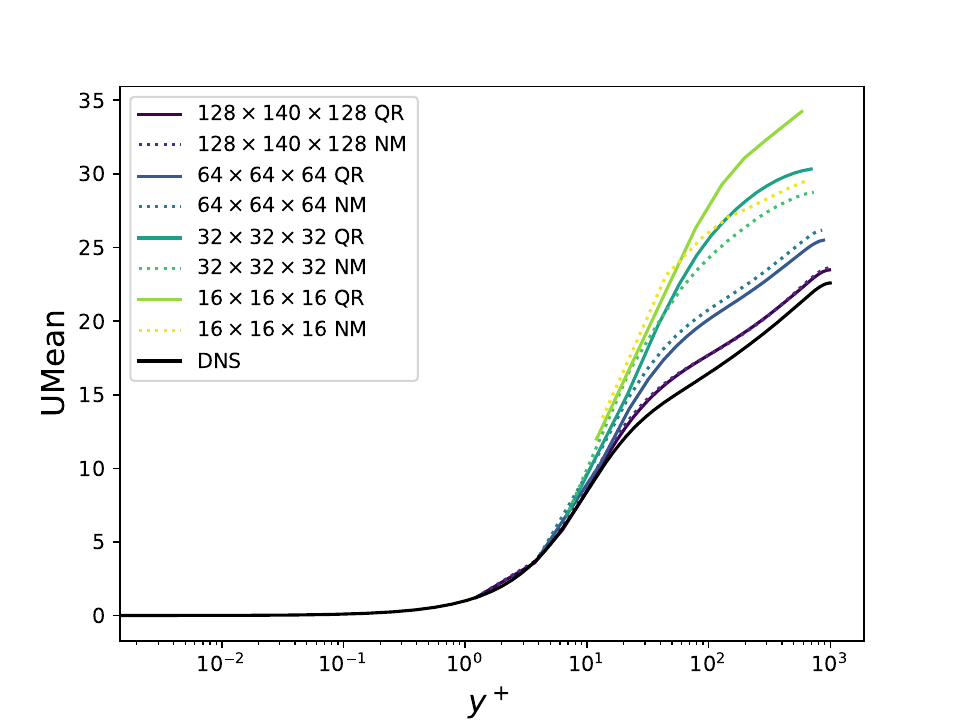}
    \includegraphics[trim = 20 0 30 25, clip,width=0.45\linewidth]{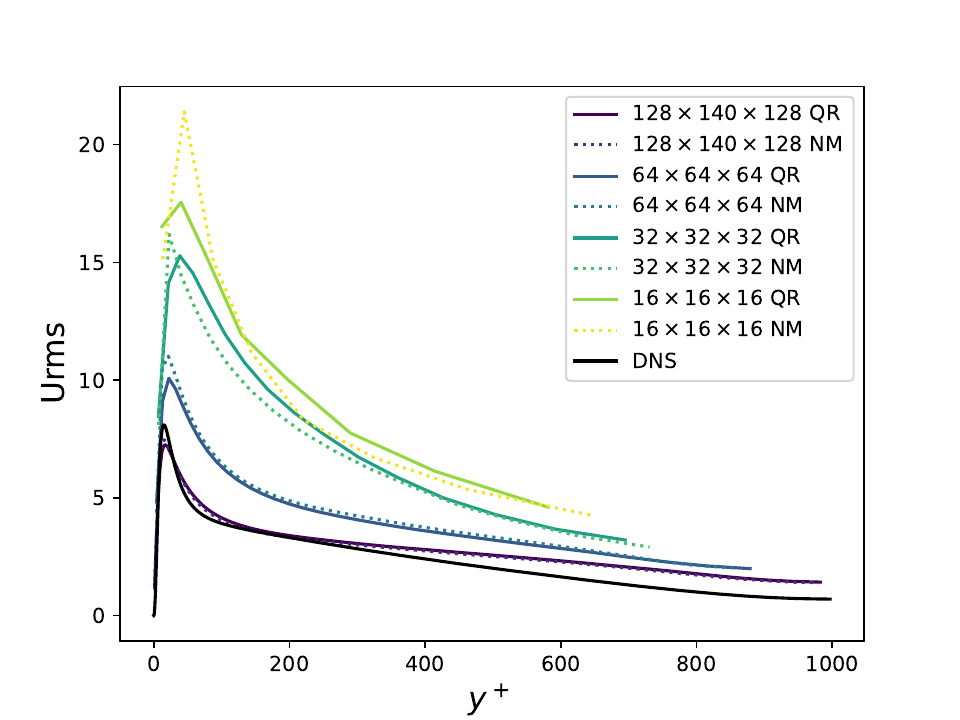}
    \includegraphics[trim = 10 0 30 25, clip, width=0.45\linewidth]{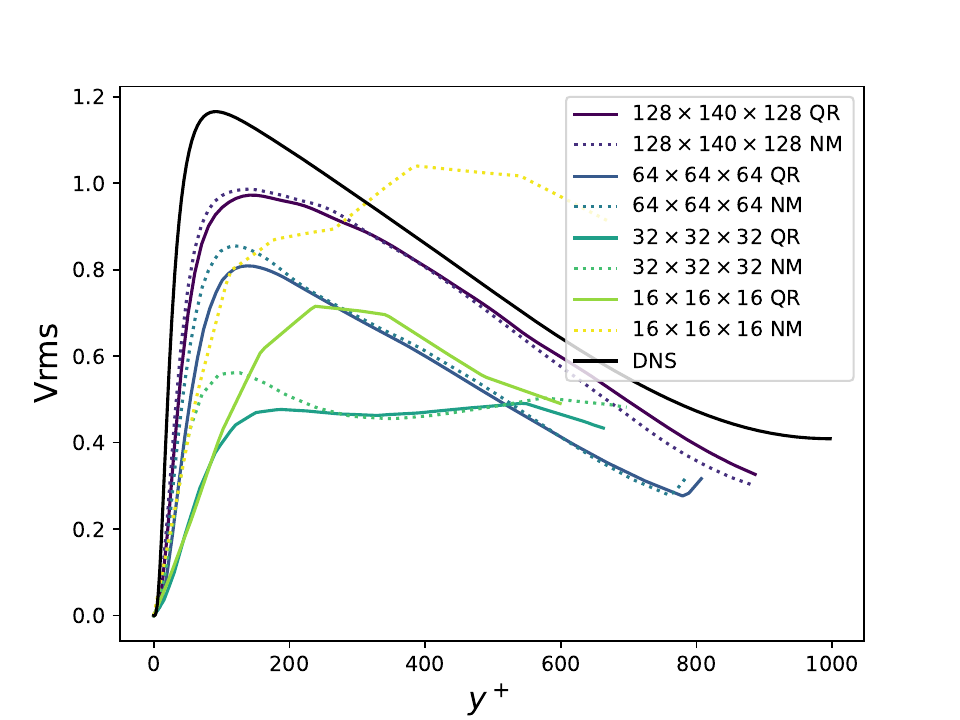}
    \includegraphics[trim = 20 0 30 25, clip,width=0.45\linewidth]{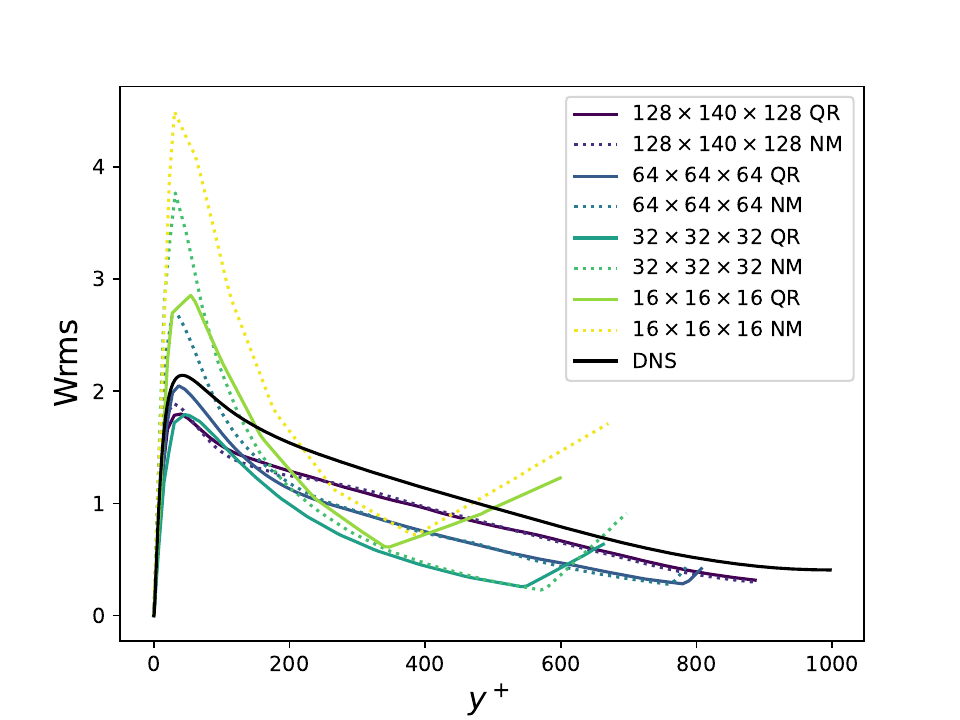}    
    \includegraphics[trim = 20 0 30 25, clip,width=0.45\linewidth]{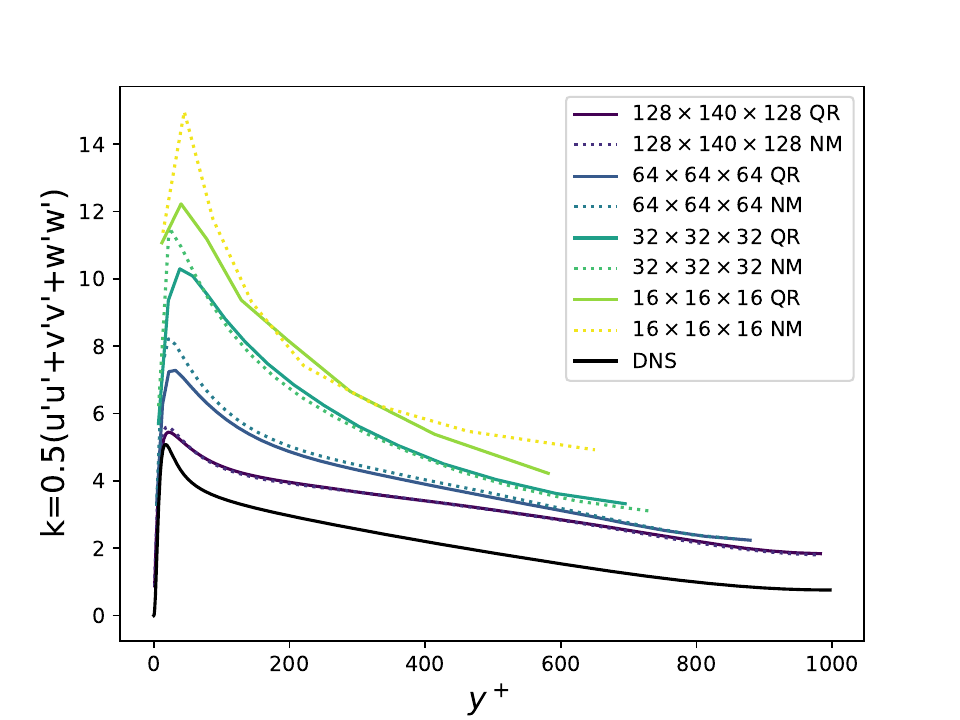} 
    \includegraphics[trim = 20 0 30 25, clip, width=0.45\linewidth]{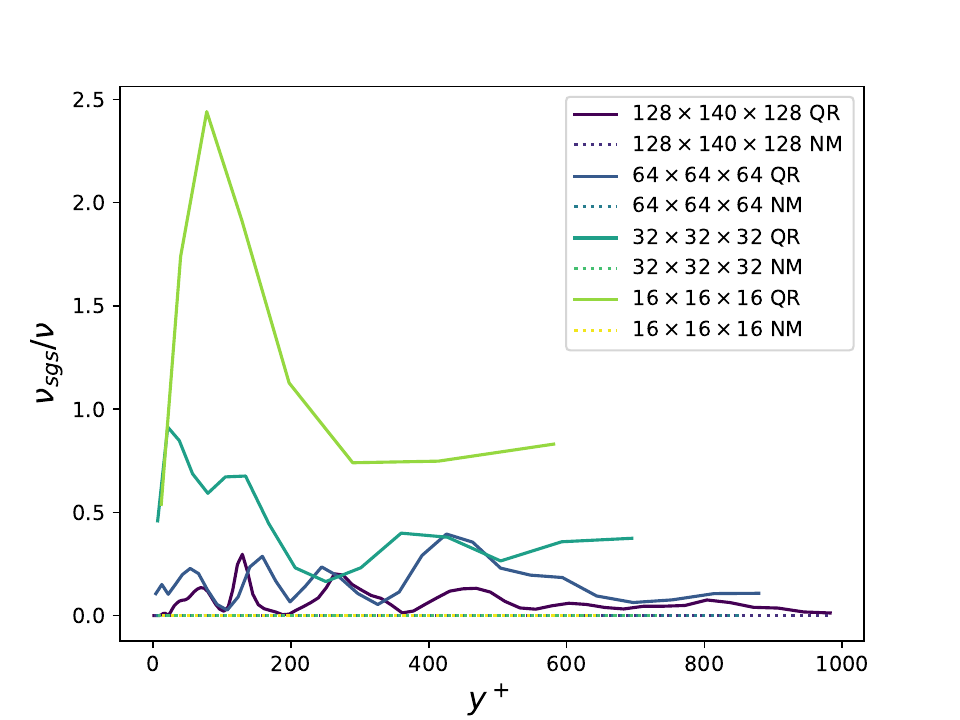}
    \caption{Comparison of reference DNS and QR simulations for the mean, RMS velocity profiles, turbulent kinetic energy, and the ratio of the sub-grid scale viscosity and molecular viscosity for fully developed turbulent channel flow at $Re_\tau$ = 1000 with four different meshes at $Re_{\tau}$ = 1000. QR denotes the QR model; NM denotes the no model simulation, i.e. the model constant is zero.}
    \label{fig:channel6}
\end{figure}
As a starting point, we present the effect of the LES model by systematically performing the computations for fully developed turbulent channel flow using four mesh resolutions as specified in Table \ref{tab:channel5}. Note the friction velocity $u_\tau$ and friction Reynold number $Re_\tau$ are decreasing with the decrease of the mesh points, this is due to the mesh size next to the wall getting smaller. The presented mean and RMS velocity and the kinetic energy profiles are normalized using the friction velocity obtained from the corresponding simulations. The main trends can be summarized as follows:

For the mean streamwise velocity $u^+$, a reduction of the grid points from $128^3$ to progressively coarser meshes results in increasingly larger over-predictions of the mean velocity for the region $y^+>10$. The QR model further overpredicts the mean streamwise velocity on $ 32^3$ and $16^3$ grids, while giving a smaller overprediction on the $64^3$ mesh compared to the no-model simulations;

For the velocity fluctuation in the streamwise direction $u'u'$, a reduction of the grid points from $128^3$ to progressively coarser meshes results in increasingly significant over-predictions of the peak in the near wall region and relatively smaller over-predictions in the channel center. By employing the QR model, the peak value of  $u'u'$ approaches the DNS data more closely in the near-wall region on all four meshes;

For the velocity fluctuation in the wall-normal direction $v'v'$, a reduction of the grid points from $128^3$ to $64^3$ results in over-predictions of the peak in the near wall region and relatively smaller under-predictions in the channel center. The coarser meshes of $32^3$ and $16^3$ failed to capture this spanwise fluctuation.

For the velocity fluctuation in the spanwise direction $w'w'$, a reduction of the grid points from $128^3$ to progressively coarser meshes results in increasingly significant over-predictions of the peak in the near wall region and relatively smaller over-predictions in the channel center. By employing the QR model, the peak value of  $w'w'$ approaches the DNS data more closely in the near-wall region on all four meshes points. Particularly, the QR model improves the prediction significantly on the coarse meshes consisting of $32^3$ and $16^3$ grid points;

For the turbulent kinetic energy $k$, a reduction of the grid points from $128^3$ to progressively coarser mesh results in increasingly significant over-predictions of the peak in the near wall region and relatively smaller over-predictions in the channel center. By employing the QR model, the peak value of $k$ approaches the DNS data more closely in the near-wall region on all four meshes;

With the decrease of the mesh points, the SGS contribution increases apparently. The contribution of the sub-grid scale model to the diffusive flux is much lower than the contribution of the molecular viscosity on fine meshes of $128^2$ and $ 64^3$ grid points.

\subsubsection{Simulation at high Reynolds number Re$_\tau$ = 1906}
Now the Reynolds number $Re_\tau=1906$ (based on channel half-width) is considered. Data from this study can be compared with the data from Moser et al\cite{lee2015direct} at $Re_{\tau}=1995$.
The physical and numerical parameters for the test cases are listed in Table \ref{tab:channel1}. The computational domain is $2\pi \times 2\delta \times \pi$  with a grid resolution of $256^3$. The normalized uniform grid spacing in the streamwise direction is $\Delta x^+=47$ and in the spanwise direction is $\Delta z^+=23$, and the first normalized wall-normal grid point next to the wall is  $\Delta y_w^+=1.8$. 

\begin{figure}[htbp!]
    \centering
    \includegraphics[width=0.49\linewidth]{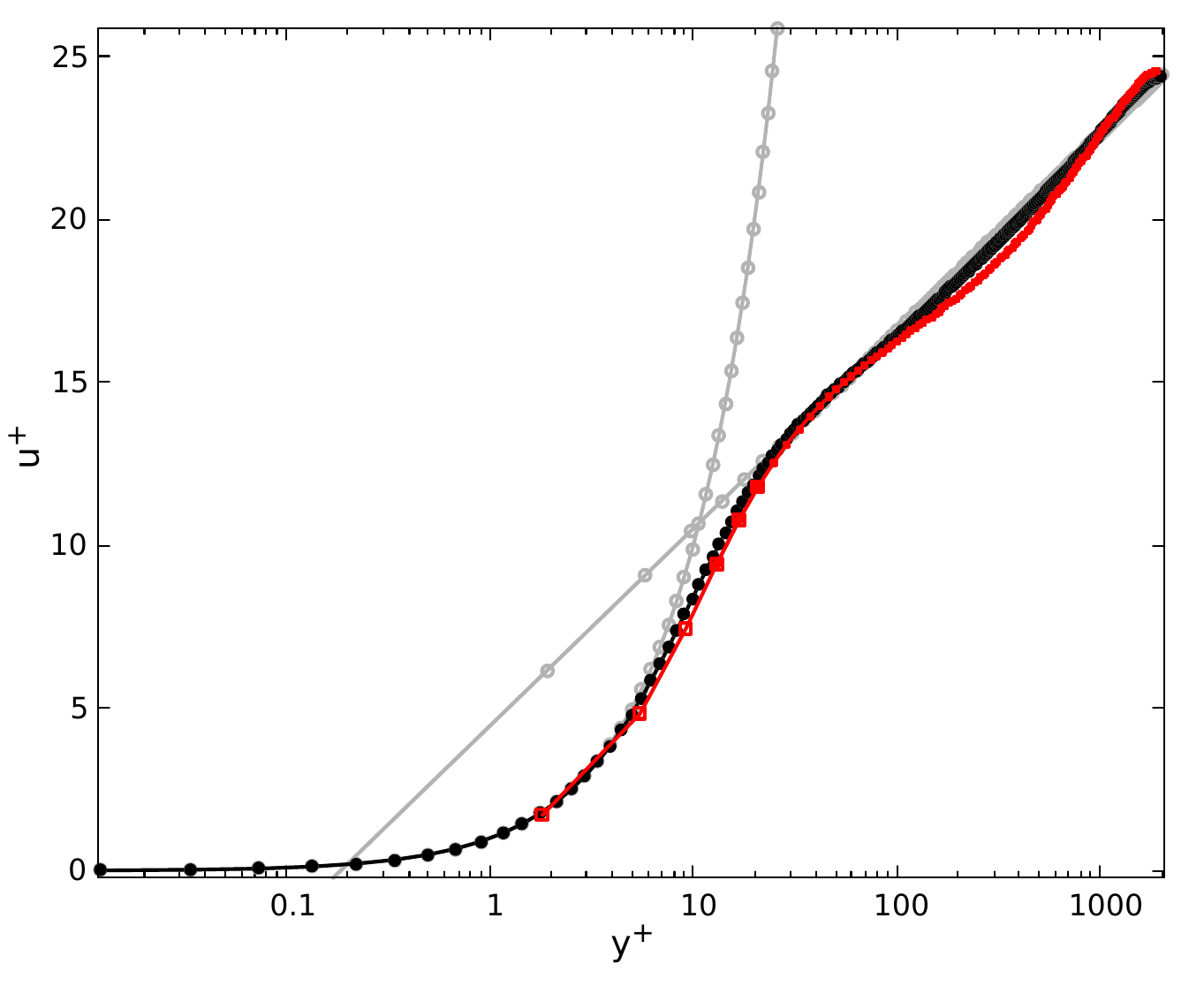}
    \includegraphics[width=0.49\linewidth]{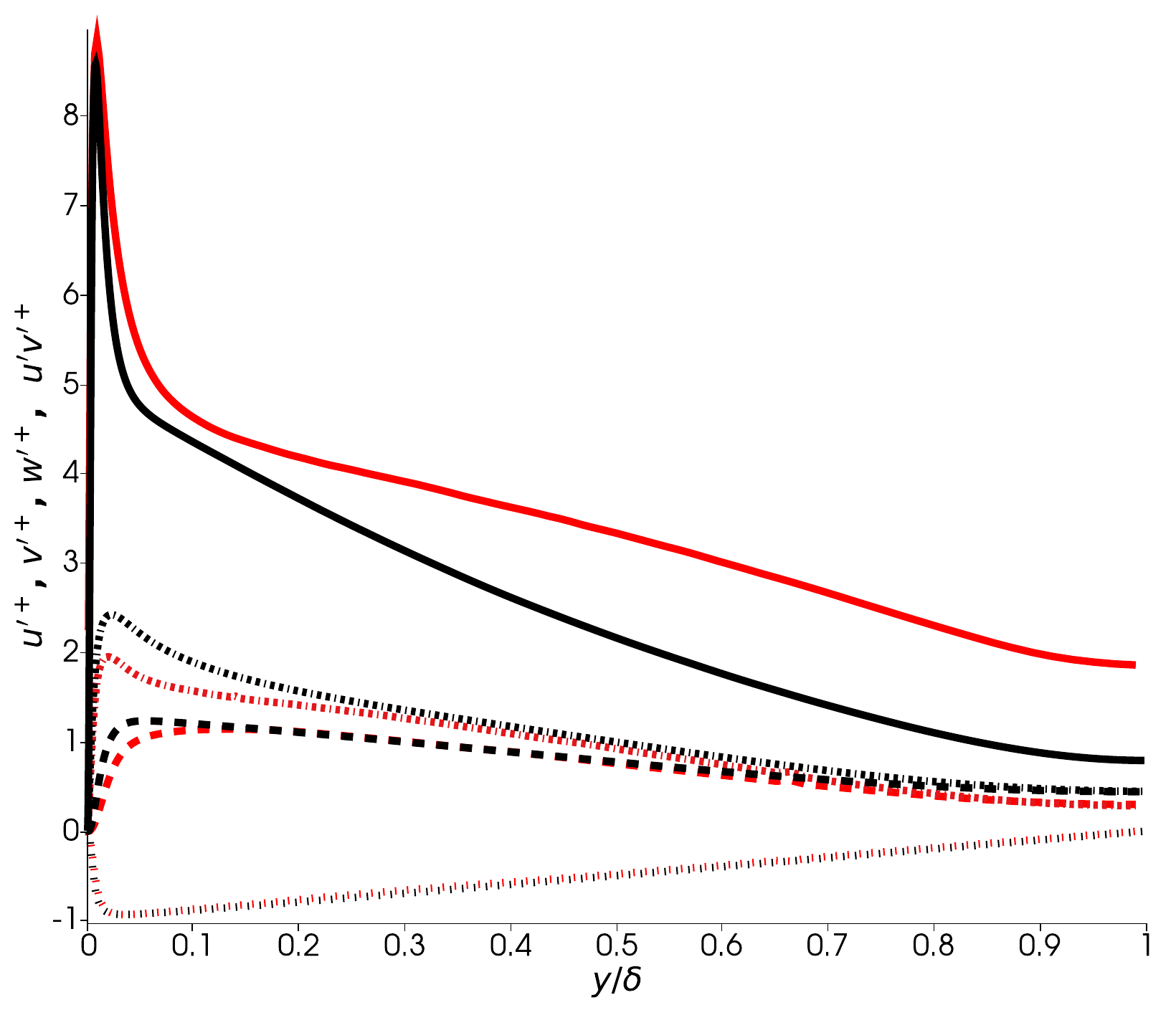}
    \captionsetup{font={footnotesize}}
    \caption{Normalized mean streamwise velocity and Reynolds stress against wall distance in wall units at $Re_\tau=1906$. Red: Minimum-dissipation model at $Re_\tau=1906$; Black: $Re_{\tau}=1995$ from Moser et al. Right: Solid line, the variance of $u$; Dash line, the variance of $v$. Dash-dot line, the variance of $w$. Dot line, covariance of $u$ and $v$.}
\label{fig:channel10}
\end{figure}

\paragraph{Mean velocity at Re$_\tau$ = 1906}
The mean velocity at $Re_\tau=1906$ is presented in the left-hand side of Figure \ref{fig:channel10}, and the Reynolds stresses are shown in the right-hand side of Figure \ref{fig:channel10}. Because the Reynolds number of the LES is larger than that of the DNS, the scaling friction velocity is adjusted so that the LES results correspond to the reference DNS data.
It is obvious that a very good agreement can be observed in the lower part of the computational domain close to the lower wall $(y^+<35)$ without any wall models.  In the region $(30<y^+<120)$, the QR results are in agreement with DNS data and log law $U^+= \frac{1}{\kappa}log{y^+} + B$, where $\kappa=0.38$ is the van Karman constant and $B=4.45$, which means there is a log-layer in this region. There is an insignificant discrepancy in the region $200<y/\delta< 500$. 

\paragraph{Reynolds stress at Re$_\tau$ = 1906}

The right-hand side of Figure \ref{fig:channel10} depicts the Reynolds stresses $\langle u'u'\rangle ^+$,$\langle v'v'\rangle ^+$, $\langle w'w'\rangle ^+$, and $\langle u'\rangle \langle v'\rangle ^+$ at $Re_\tau=1906$. 
To start the streamwise Reynolds stress is considered. It is found that $\langle u'u'\rangle ^+$ accurately matches the DNS results in the very near-wall region.  In accordance with the DNS results, present studies demonstrate that the peak value of $\langle u'u'\rangle ^+$ appears at $y/\delta \approx 0.013$.

In addition, it is quite revealing that the shear stress $\langle u'\rangle \langle v'\rangle ^+$ is in agreement with Moser's findings. The data of the variance of $v$ illustrates that the values of $\langle v'\rangle \langle v'\rangle ^+$ are in agreement with Moser's data. Besides, in the outer layer ($0.2<y/\delta<0.6$), the variance of $w$ is in line with those of DNS. 

The disagreements compared to DNS data are listed here. Adjacent to the wall, the peak value of the variance of $u$ has 4$\%$ error. Close to the center of the channel ($0.8<y/\delta<1$), the discrepancy is becoming larger. On top of that, there is also a large discrepancy in the peak value of the variance of $w$, the possible reason could be the interpolation error introduced in post-processing. There are only 4 sample points in the region $y/\delta<0.1$ (ten times less than DNS). 
There are, however, other possible explanations for these results given in the following section \ref{Explanation of the discrepancy}.

\subsubsection{Explanation of the discrepancy} \label{Explanation of the discrepancy}
With respect to the disagreements in the results at the high Reynolds number $Re_\tau=1006$ and $1906$, there are several possible explanations. First of all, the difference between the Reynolds number in QR simulation and reference data is the critical factor that causes the mismatch in all the comparisons. 

Besides, the discrepancy could be attributed to the small domain size of LES (four and three times smaller in the streamwise direction and spanwise directions respectively), as reported in Lozano-Durán and Jiménez's work \cite{lozano2014effect}. Moreover, it is very likely that the QR simulations have inadequate mesh points. The mesh resolution of QR is eighteen, four, and sixteen times coarser in streamwise, wall-normal, and spanwise directions, respectively. It has been proved that by increasing the mesh points from 128 to 140 in the spanwise direction, the result already got improved. Further refinement is computation cost and out of the scope of our interest. Last but not least, the numeric schemes in OpenFOAM are only second-order accurate. Compared to seventh-order accurate schemes in Moser's data, the effects of numerical dissipation in LES simulation are important.

In conclusion, the various comparisons carried out have demonstrated that the QR prediction is reliable. A properly chosen value of the QR coefficient C = 0.024 can provide very similar results as a dynamic model but with a reduced computational cost. The near wall region is captured accurately with any wall functions. The findings obtained at high Reynolds numbers mainly confirmed the accuracy of predicting turbulence with relatively coarse mesh. The contribution of the sub-grid scale model increases with the decrease of the mesh points, and the eddy-dissipation is much less than the molecular dissipation in the simulation at fine meshes. The symmetry-preserving discretization combined with the QR model outperforms the standard second-order accurate discretization method in OpenFOAM.

\section{Flow over periodic hills}
Flow separation from curved surfaces and subsequent reattachment is a flow phenomenon often appearing in engineering applications.  To assess the applicability of the proposed minimum-dissipation model in OpenFOAM to compute separated flows, simulations of three-dimensional flow over periodic hills at $Re=10595$ have been performed.

The geometry retains the shape of the hill defined by Mellen et al.\cite{mellen2000large}.  The hill height is $H=28mm$, and hill crests are separated by $L_x=9H$. The distance between two consecutive hills is $x/H = 5.142 $ to enhance the streamwise decorrelation. This configuration facilitates the natural reattachment of the flow between two successive hills and establishes a considerable post-reattachment-recovery region on the flat plate between hills prior to the flow's re-acceleration over the subsequent hill. The channel height and spanwise width are $L_y = 3.035H$ and $L_z=4.5H$, giving the aspect ratio $L_z/L_y$ =1.483. The Reynolds number $Re=10594$ is based on the hill height $H$, the bulk velocity $U_b$ taken at the crest of the first hill and the kinematic viscosity $\nu $ of the fluid.

The flow is assumed to be periodic in the streamwise direction and thus periodic boundary conditions are applied.
In analogy to the turbulent plane channel flow case, the non-periodic behavior of the pressure distribution can be accounted for by adding the mean pressure gradient as a source term to the momentum equation in the streamwise direction. To ensure a fixed Reynolds number, the chosen approach is to maintain constant mass flux, which necessitates the adjustment of the mean pressure gradient over time.
Additionally, the flow is assumed to be homogeneous in the spanwise direction, and periodic boundary conditions are implemented accordingly. The simulations are conducted on a grid consisting of approximately 2.56 million points. The grid resolution near the wall is sufficient to resolve the viscous sublayer, as indicated by a $y^+$ value of approximately 0.17 at the closest grid points to the wall. Therefore, the no-slip boundary condition is employed at the wall.

For the initialization of the transient state, the RANS simulation is performed using the Spalart-Allmaras model. In order to minimize statistical errors resulting from insufficient sampling, the flow field is averaged in the spanwise direction and over an extended period of time. The averaging period covers a time interval of approximately 40 flow-through times.

\begin{figure}[ht!]
\centerline{
 \includegraphics[trim=50 80 40 90, clip, width=0.5\linewidth]{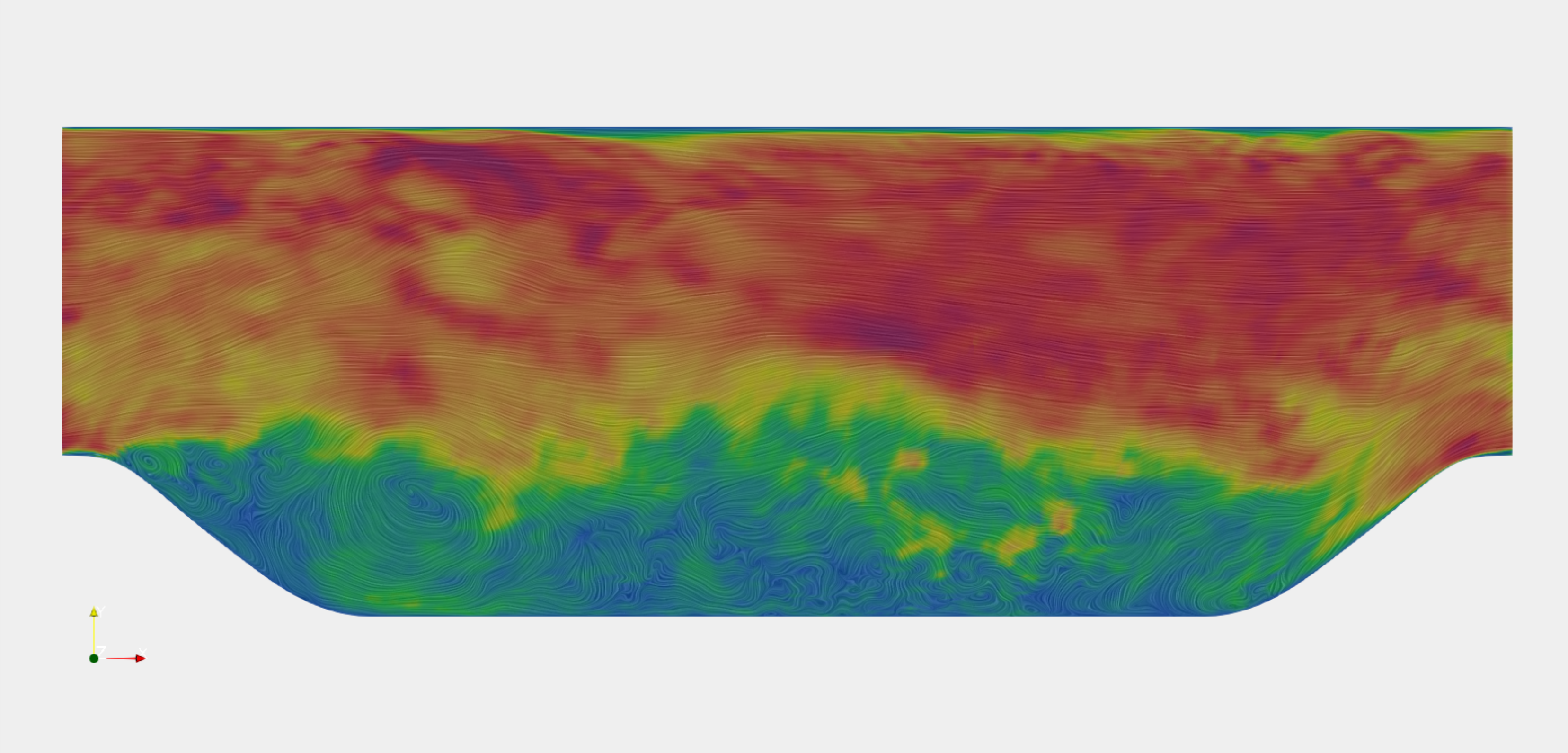}
 \includegraphics[trim=40 80 50 90, clip, width=0.5\linewidth]{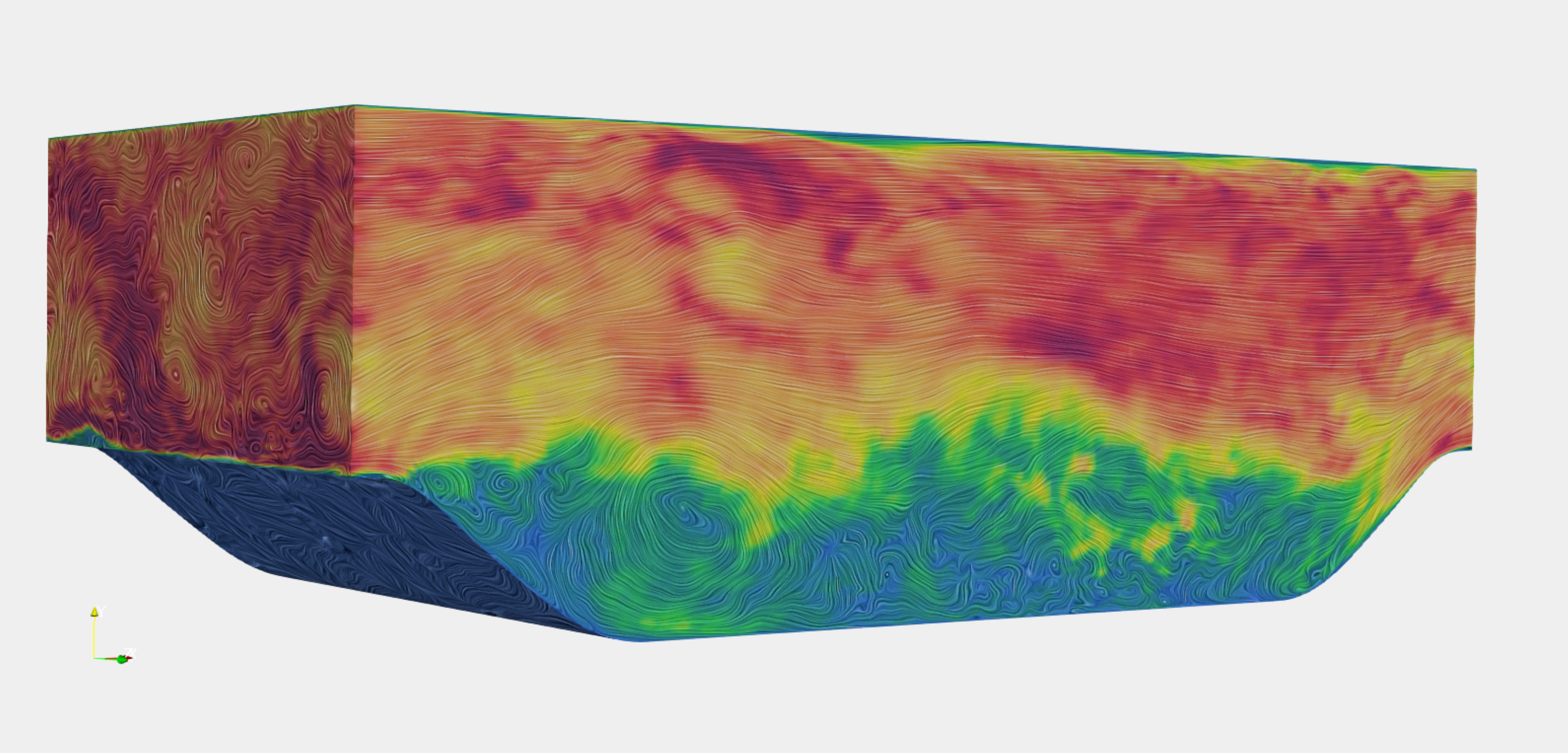}}
 \captionsetup{font=footnotesize}
\caption{The geometry of the periodic hills. Left: the front view, right: the side view}
\label{fig:hill1}
\end{figure}

\subsection{Cross-Comparison of calculation from QR model using standard OpenFOAM discretization with literature data}

This investigation focuses on the physical aspects of the flow considered. The flow over periodically arranged hills separates from a curved surface, recirculates on the leeward side of the hill, and reattaches naturally at the flat channel bottom.   
The location $x/H=0.05$ is at the narrowest cross-section, here the most intensive acceleration occurs locally and globally.  The position $x/H=0.5$ is located shortly after the separation line and crosses the strong shear layer; the profile of $x/H=2$ corresponds to the beginning of the flat floor and hence within the main recirculation region; $x/H=4$ is located near the end of the recirculation bubble and finally, $x/H$=6 is positioned behind the main separation region in the reattached flow.

\subsubsection{Separation and Reattachment Lengths}
The separation and reattachment points are obtained at $Re=11230$, i.e. the bulk velocity $u_b=1.06$, using Gauss linear spatial discretization and Backward temporal discretization. The separation point is accurately determined by numerically solving the boundary layer equations under pressure-adverse conditions. The QR model predicts the separation point, where the wall shear stress reaches zero, to be approximate $x/H \approx 0.175$, which is smaller than the reference value of $x/H \approx 0.19$.
This discrepancy is reasonable since the separation point moves upstream with increasing Reynolds numbers. The separation point has a strong impact on the point of reattachment. The recirculation starts at $x/H \approx 0.27$ and ends on $x/H \approx 4.71$. The length of the main recirculation bubble is approximately 4.48. The reattachment position where the dividing streamline attaches to the wall again is $x/H = 4.71$, which is very close to the reference value of 4.69 \cite{rapp2010new}. 

\paragraph{The effect of Reynolds number}
The mean velocity and Reynolds stress predicted by the QR model without a wall damping function are compared with experimental data \cite{rapp2010new} and three CFD tests \cite{temmerman2001large}. For this comparison, one should bear in mind that all LES simulations are based on second-order accurate numerical techniques in space and time. Besides, the grid of the QR model consists of six times fewer grid points than the grid that used in reference CFD data. 
The QR simulation mesh consists of 2.56 million grid points and requires six hours to simulate on two nodes with 128 cores each. This mesh is sufficient for modeling this case, as further refinement does not yield improved results.

\begin{figure}[ht]
    \centering
            \hspace{-1.55cm}\includegraphics[width=0.99\linewidth]{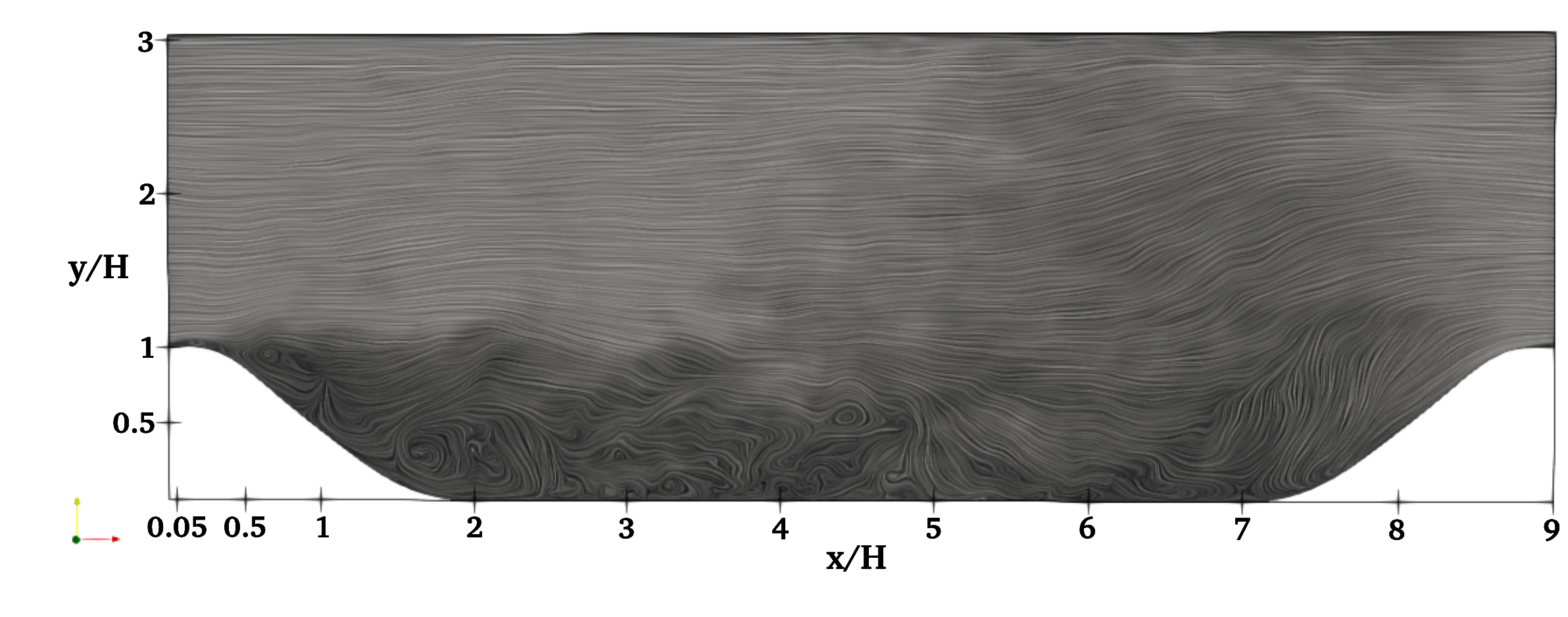}
            \vskip -6.1cm
            \includegraphics[width=0.91\linewidth]{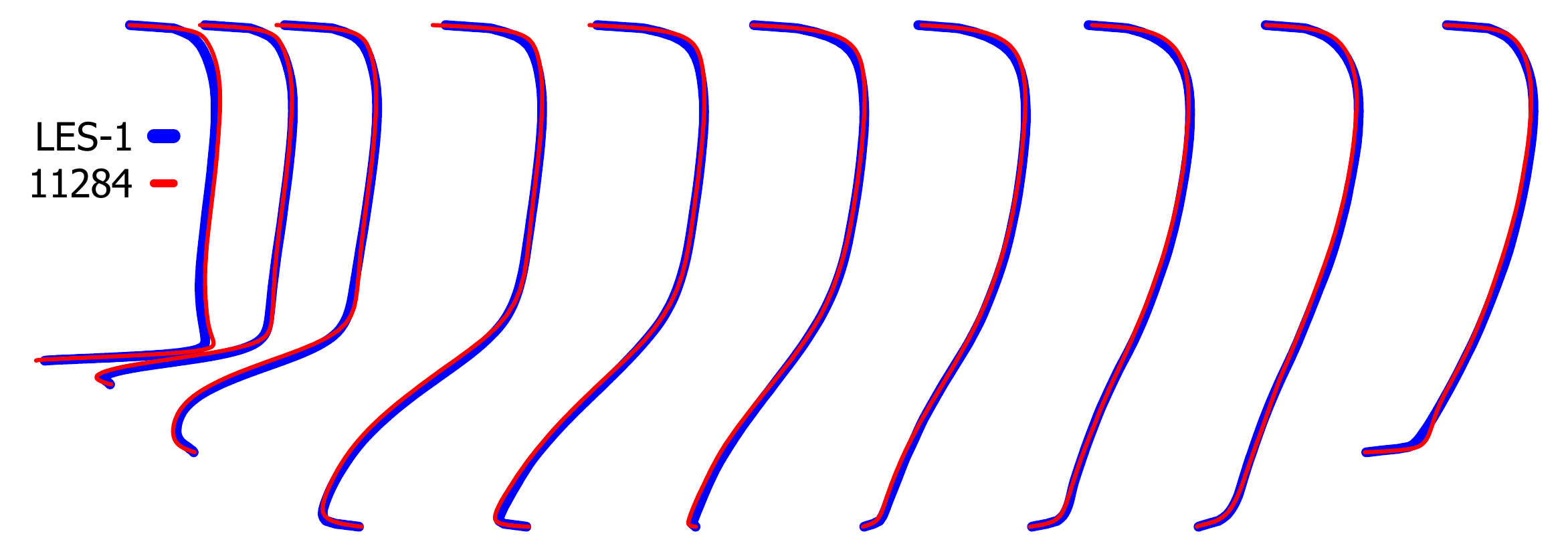}
            \vskip 0.51cm
            \includegraphics[width=0.96\linewidth]{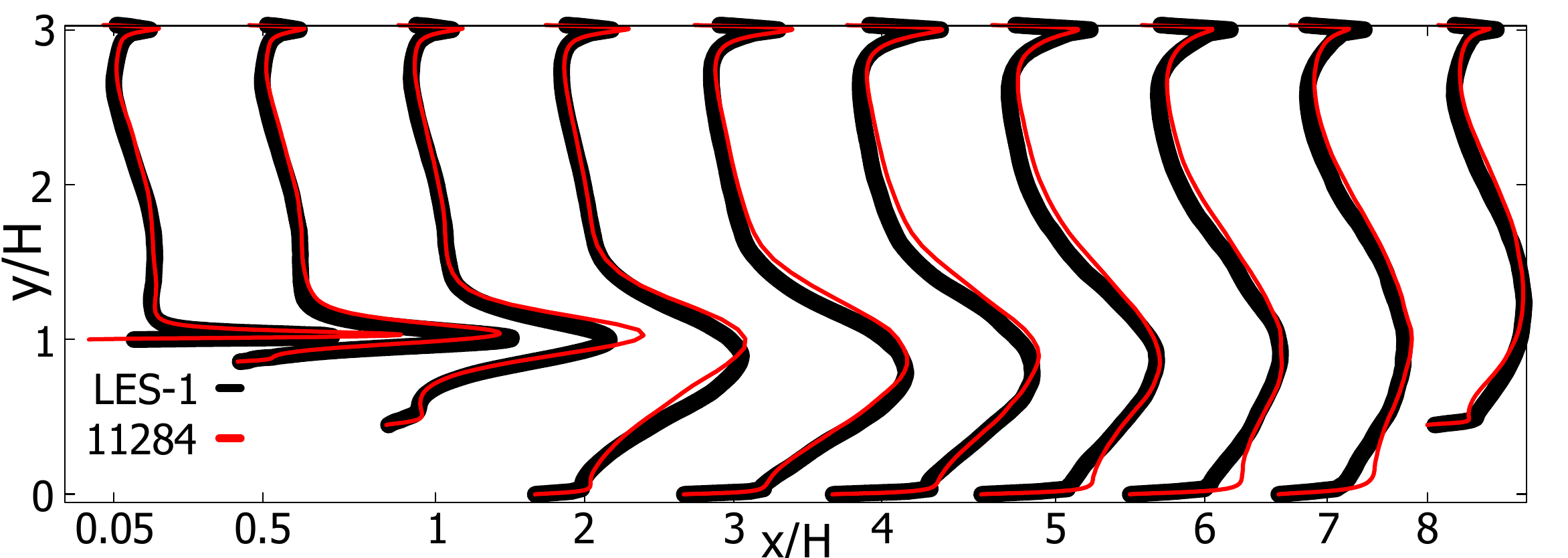}
            \captionsetup{font={footnotesize}}
            \caption{Streamwise velocity (top) and Reynolds stress (bottom) at ten different positions with the increment of Reynolds number comparing with large-eddy simulation data}
            \label{fig:hill4}
\end{figure}

To mitigate the underprediction of streamwise velocity component $\langle u\rangle /u_b$ in $1.5< y/H < 2.8$ while calculating with Reynolds number the same as reference $Re=10,595$, the simulations with higher Reynolds numbers have been investigated. The most striking result to emerge from the data, as indicated in Figure  \ref{fig:hill4}, is that the mean velocity and Reynolds stress obtained with the 4$\%$ larger Reynolds number is in agreement with the B-spline reference. In addition, a 7$\%$ larger Reynolds number is consistent with the experimental data. These findings confirmed that increasing the Reynolds number by $3\%- 7\%$ can match the reference data.  Granted that the experiment and three LES simulations used to compare were performed at the same Reynolds number $Re=10595$, they do not fully agree with each other.

\subsection{Symmetry-preserving discretization compared to the standard OpenFOAM discretization schemes}
In this section, the symmetry-preserving discretization implemented in OpenFOAM is compared with the standard Gauss linear schemes in OpenFOAM and the experimental data obtained from Temmerman and Leschziner\cite{temmerman2001large}. 
We provide a list of the common parameters employed in both Large-Eddy Simulations. The QR model with a model constant of $C=0.024$ is utilized as the large-eddy model. The bulk velocities in two LES simulations are $U_b=1.05$ $m/s$. The fluid viscosity is $\nu=$ 2.643 $\times 10^{-6} $ $m^2/s$. There is no wall function applied in all the simulations. The time-step is $\Delta t=1\times 10^{-4}s$ and the maximum CFL number is limited to 0.78 to ensure numerical stability. The size of the computational domain, mesh resolution, initialization, and post-processing are the same as in previous simulations. 

To ensure comparability between the two cases, we have chosen second-order accurate numerical schemes. In the standard OpenFOAM simulations, the following numerical schemes are applied. The temporal discretization is performed using the implicit backward scheme. The gradient, divergence, and Laplacian terms are discretized using the Gauss linear (central difference) schemes. The pimpleFoam solver is employed to solve the governing equations. The pressure equation is solved using the GAMG (geometric agglomerated algebraic multi-grid) solver with DICGauss-Seidel (diagonal incomplete-Cholesky with Gauss-Seidel) smoother.The velocity equation is solved using the PBiCGStad (stabilized preconditioner bi-conjugate gradient for both symmetric and asymmetric matrices) with DILUpreconditioner (simplified diagonal incomplete LU preconditioner for asymmetric matrices). The number of outer correctors for performing the momentum equation is set to 10, the number of inner correctors for correcting the pressure within an iteration is set to 2 (suggested to be 1-3 in the PIMPLE guide), and the number of non-orthogonal correctors is 1. 

The symmetry-preserving discretization is described in section \ref{sec:symmetry}. The solver is called RKSymFoam. The temporal discretization is performed using the implicit Crank-Nicolson scheme.   The pressure equation is solved using the GAMG solver with DICGauss-Seidel smoother. The velocity equation is solved using the PBiCGStad (stabilized preconditioner bi-conjugate gradient for both symmetric and asymmetric matrices) with DILUpreconditioner (simplified diagonal incomplete LU preconditioner for asymmetric matrices).
The outer corrector for updating the non-linear convective term is 10, the inner PISO iteration loop for the pressure-velocity coupling is 2, and the number of non-orthogonal correctors is 1. 

\begin{figure}
    \centering
    \includegraphics[width=0.96\linewidth]{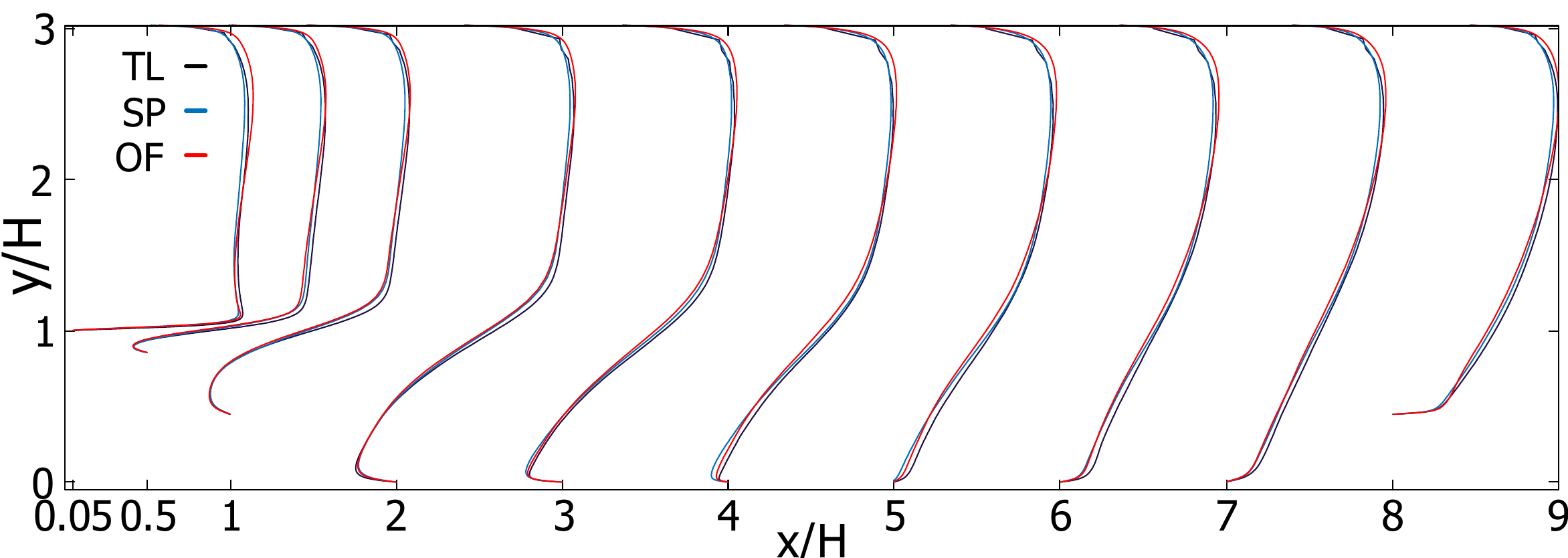}
    \includegraphics[width=0.96\linewidth]{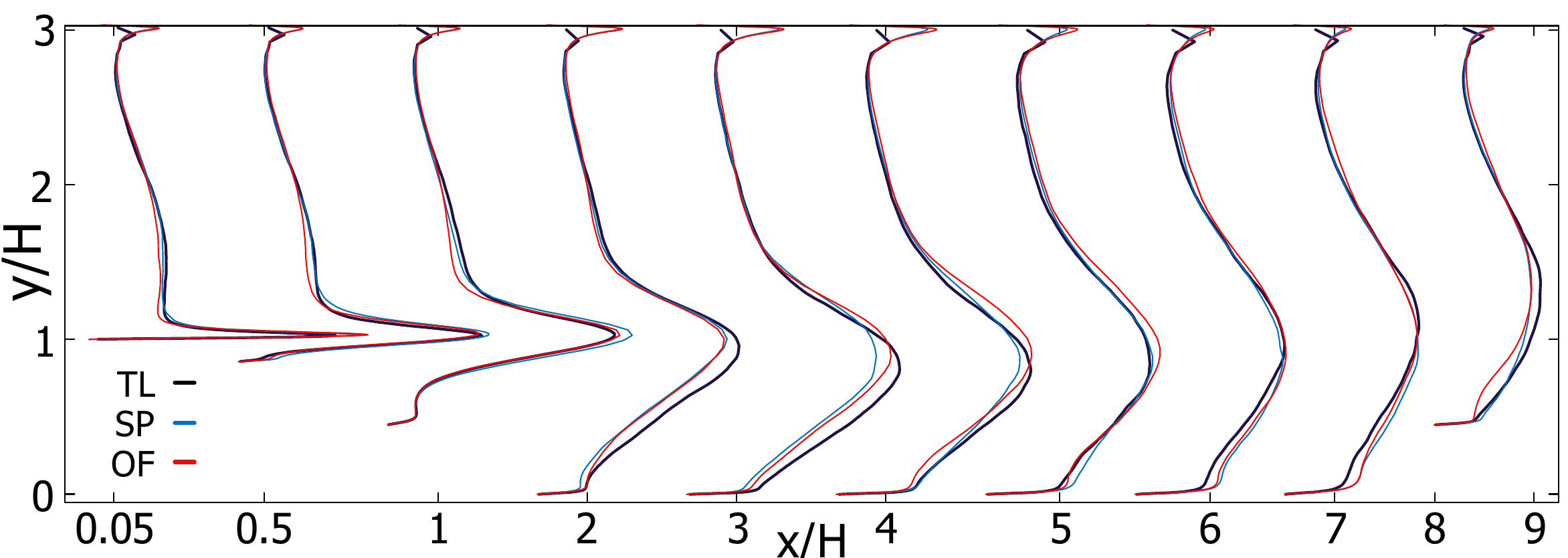}
    \caption{Comparison of the predictions by standard OpenFOAM discretization and symmetry-preserving discretization. Top: Mean streamwise velocity; middle: Averaged Reynolds stress in the streamwise direction $u'u'$ at ten different locations in the streamwise direction; Bottom left: Zoomed-in $u'u'$ at $x/H=0.05, x/H=0.5$ and $x/H=1$; (d): Eddy viscosity normalized by the fluid viscosity.}
    \label{fig:hill7}
\end{figure}
 The separation point predicted by the symmetry-preserving discretization is approximately $x/H \approx 0.175$, which is smaller than the reference value of $x/H \approx 0.19$. The recirculation starts at $x/H \approx 0.27$ and ends on $x/H \approx 5.02$. The length of the recirculation bubble is approximately $x/H\approx 4.7$. 

As we can see from Fig.\ref{fig:hill7}, the mean velocity predicted by the two discretization schemes is consistent in the upper part ($y/H> 1$) of the computational domain, where the structure is relatively simple, and no hill is present. On the bottom part ($y/H <1$), two simulation results are again similar in the upstream region $x/H=0.05, 0.5, 1$ and $2$. However, from $x/H=3$ onwards until the end of the domain, both simulations underpredict the velocity, with the symmetry-preserving schemes exhibiting a greater underprediction compared to the standard OpenFOAM scheme. Notably, in the channel flow simulation at $Re_\tau =1000$, the symmetry-preserving discretization proves to be more accurate than the central difference schemes employed in OpenFOAM. This superiority of symmetry-preserving discretization has also been found at $Re_\tau =180$ \cite{komen2021symmetry}.

The Reynolds stress in the streamwise direction $u'u'$ in the middle of Fig.\ref{fig:hill7} shows the underprediction and overprediction at different locations. The trends are clearer if the upstream region is zoomed in, as shown in the bottom figure. In the first three locations, i.e. $x/H=0.05, 0.5$, and $1$, the standard OpenFOAM underestimate the $u'u'$ at the middle ($1<y/H<2$), overestimates the peak value ($y/H=1$) and the $u'u'$ near the wall. Meaning the acceleration predicted by standard OpenFOAM is more intense in the shear layer at the hill crest. Fig.\ref{fig:hill7}(d) shows the eddy viscosity normalized by the fluid viscosity in the spanwise direction, from which we can see the model contribution $\nu_t/\nu$ is below 0.3.

To sum up, for simulating periodic hills, the minimum-dissipation model, along with standard OpenFOAM discretization schemes and symmetry-preserving schemes, provides dependable results while significantly reducing computational expenses. The symmetry-preserving discretization yields more accurate outcomes in certain areas of the computational domain. Komen et al.\cite{komen2021symmetry} found that the numerical dissipation introduced by standard OpenFOAM discretization exceeds the contribution of the large-eddy model. Therefore, combining the symmetry-preserving discretization with the QR model is advantageous and dependable.

\section{Flow over circular cylinder}
\subsection{Numerical Method}
\begin{figure}[ht!]
\begin{subfigure}[b]{0.49\textwidth}
    \centering
    \includegraphics[trim=200 50 250 80, clip, height=0.19\paperheight]{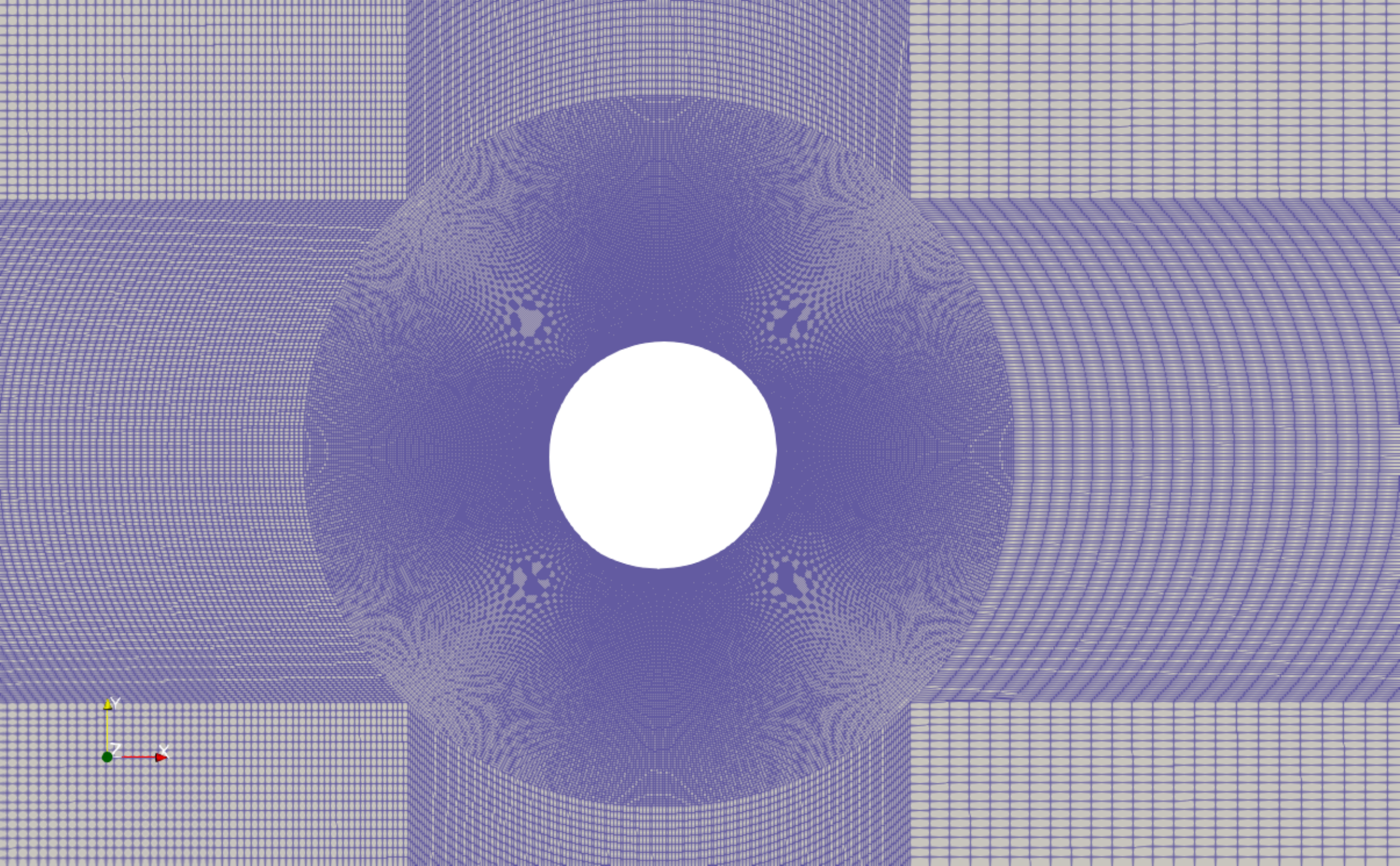}
    \captionsetup{font={footnotesize}}
    \caption{The magnified O-mesh in the vicinity of the cylinder}
\end{subfigure}
\begin{subfigure}[b]{0.49\textwidth}
    \centering
    \includegraphics[trim=0 230 0 100, clip, height=0.094\paperheight]{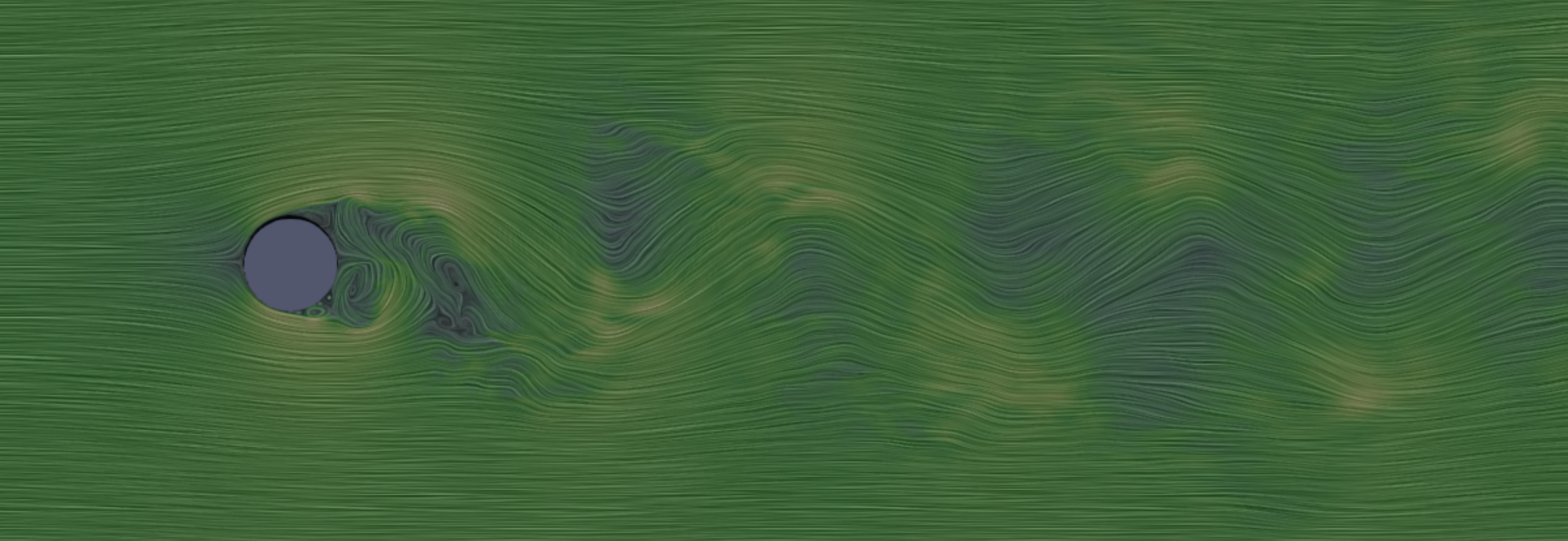}
    \includegraphics[trim=0 180 0 150, clip, height=0.094\paperheight]{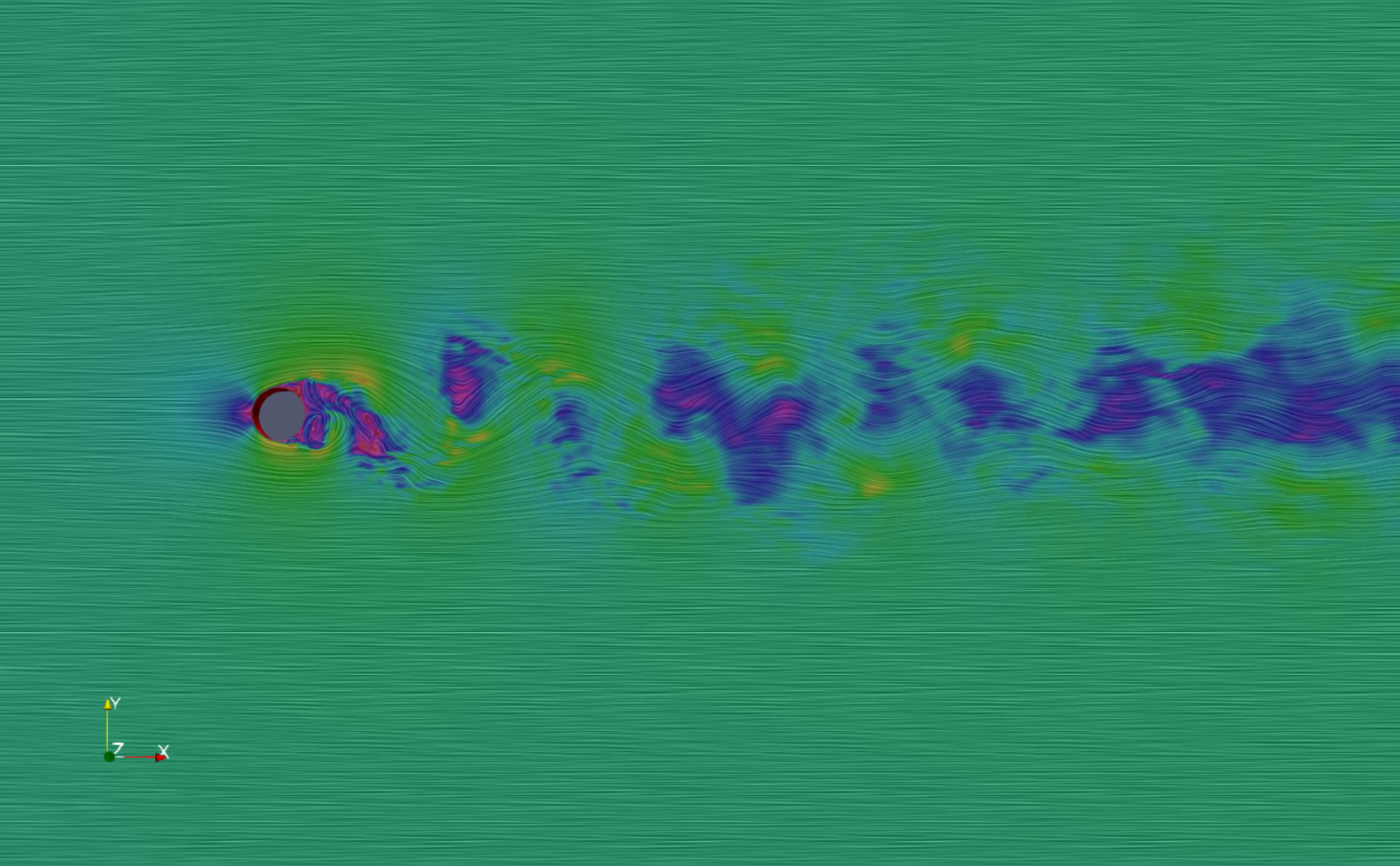}
    \captionsetup{font={footnotesize}}
    \caption{The instantaneous velocity in computational domain}
\end{subfigure}
\caption{The flow over cylinder at $Re_D=3900$ with $50D \times 30D \times \pi D$ of  the domain size}
\label{fig:cylinder1}
\end{figure}

The computational geometry of the region of interest is shown in Fig  \ref{fig:cylinder1}. A cylinder (diameter D = 1m) is placed at 10D from the inlet and at 40D from the outlet in the domain, which has size $50D \times 30D \times \pi D$. The periodicity is imposed in the spanwise direction of the circular cylinder. The constant free-stream velocity $U = 1 m/s$ is used to describe the inlet flow. The zero gradient condition is adopted for the outflow. In the plane normal to the cylinder axis, an O-type mesh is adopted. The simulation is performed in OpenFOAM using a PISO loop to solve the governing equations. The temporal discretization is performed with Euler, backward and Crank-Nicolson schemes. The details of numerical parameters are listed in Table \ref{tab:cylinder1}.

\begin{table}[h]
    \small
    \centering
    \begin{tabular}{ccccccc}
    \hline
 case   &   time scheme &     div  	&	grad(U)   &   $N_{tot}$    &   $N_{z}$   & $L_x \times L_y \times L_z$\\ \hline
Run \RomanNumeralCaps{1}   &	Euler	   &    Filtered  &	cellMDLimited linear  1	& 768 000 &  16  &   $50D \times 30D \times \pi D$\\  
Run \RomanNumeralCaps{2}   &    Euler      &    Filtered  &	cellMDLimited linear  1	&   6 144 000 &  32  &   $50D \times 30D \times \pi D$\\
Run \RomanNumeralCaps{3}   &   Backward    &    Linear    &   Gauss linear        &   6 144 000 &  32  &   $50D \times 30D \times \pi D$\\
Run \RomanNumeralCaps{4}   &   CN 0.9      &    LUST      &   cellMDLimited linear 1  &   6 144 000 &  32   &   $50D \times 30D \times \pi D$\\
Run \RomanNumeralCaps{5}   &   Backward    &    Linear    &   Gauss linear      &  10 086 912    &  32  &   $50D \times 30D \times \pi D$\\
Run \RomanNumeralCaps{6}   &    Euler      &    Filtered  &  cellMDLimited linear  1 &  12 288 000    &  64  &   $50D \times 30D \times 2\pi D$\\
Run \RomanNumeralCaps{7}   &	Euler	   &    Filtered  &	cellMDLimited linear  1	& 457 600 &  16   &   $50D \times 30D \times \pi D$\\ 
Run \RomanNumeralCaps{8}   &	Euler	   &    Filtered  &	cellMDLimited linear  1	& 320 000 &   16 &   $50D \times 30D \times \pi D$\\ 
Run \RomanNumeralCaps{9}   &	Euler	   &    Filtered  &	cellMDLimited linear  1	& 204 800 &  16  &   $50D \times 30D \times \pi D$\\ 
\hline
    \end{tabular}
    \captionsetup{font={footnotesize}}
    \caption{The computational parameters of flow over cylinder at $Re_{D}=3900$. $N_{tot}$: the total number of grid points, Gauss: the standard Gaussian finite volume integration, div: the interpolation scheme of divergence terms, grad(U): the interpolation scheme adopted to solve the velocity gradient term in NS equations, CN: Crank-Nicson time discretization, LUST: fixed blended scheme with 0.25-second order upwind and 0.75 central difference weights, linear: central difference interpolation scheme, Filtered: central difference with filtering for high-frequency ringing. (details can be found in the OpenFOAM documentary)}
    \label{tab:cylinder1}
\end{table}

The mean velocity and Reynolds stress from the simulations are compared with the experimental results from Lourenco and Shih\cite{lourenco1994characteristics}, Ong and Wallace\cite{ong1996velocity} and the LES results from Kravchenko and Moin\cite{kravchenko2000numerical}, Mittal and Moin\cite{mittal1995large}, Breuer\cite{breuer1998large} and Beaudan\cite{beaudan1995numerical}.
In general, statistics are compiled over twenty vortex shedding cycles or at least over a period of $T=100D/U_\infty$ to ensure the convergence of statistics. Note to capture the low-frequency component, the lift and drag coefficients, the averaging is to be done over long time periods. The averaging is also performed over the spanwise direction.

\subsection{Simulation Results and Discussion}

\subsubsection{Global parameters}
Some of the important flow parameters from our simulations are summarized in Table \ref{tab:cylinder2}. Also tabulated for direct comparison are the corresponding experimental results and three-dimensional LES simulations from various studies. Next to the large eddy simulation from Kravchenko et al.\cite{kravchenko2000numerical}, the LES results of Breuer\cite{breuer1998large} are also listed in Table \ref{tab:cylinder2}. The experiments of Lourenco and Shih did not provide values for the mean drag coefficient, and we obtained these values from the other experimental studies listed in Table \ref{tab:cylinder2}. The mean drag coefficient, root mean square of lift coefficient, recirculation length, and Strouhal shedding frequency are found to be in fairly good agreement with the results of the three previous simulations and the experimental data. Especially, the recirculation length found in our simulation with filtered central difference scheme (Run \RomanNumeralCaps{2}) is precisely in accord with the experimental data from Lourenco\cite{lourenco1994characteristics}. And the mean drag coefficient from the upwind biased scheme agrees with the experiments from Norberg\cite{kravchenko2000numerical}.

\begin{table}[htbp!]
    \centering
    \small 
    \begin{tabular}{cccccc}
    \hline
      Case  &   $\bar C_d$  &     $Cl$      &   $St$    &  $\overline{L_r}/D$\\ \hline
    Run \RomanNumeralCaps{1}   &      1.36     &    0.448   &   0.195    &      0.994       \\
    Run \RomanNumeralCaps{2}   &      1.18     &    0.286   &   0.131    &      1.189             \\
    Run \RomanNumeralCaps{3}   &      1.205    &    0.366   &   0.207    &      0.994        \\
    Run \RomanNumeralCaps{4}   &      0.982    &    0.086   &   0.214    &      1.720             \\
    Run \RomanNumeralCaps{5}   &      1.214    &    0.3994  &   0.208    &      -             \\
    Run \RomanNumeralCaps{6}   &      2.2347   &    0.599  &   0.2066    &      -             \\
      LES-C2\cite{breuer1998large} & 1.10  & -     &     -     &     1.115       \\
      LES-C3\cite{breuer1998large} & 1.07  & -     &     -     &     1.197        \\
      LES \cite{kravchenko2000numerical} & 1.04  & -   &    0.21   &      1.35       \\
   \multirow{1}{*}{Exp} & \multicolumn{1}{l}{0.99$\pm$0.05\cite{kravchenko2000numerical}} & \multicolumn{1}{l}{} &   \multicolumn{1}{l}{0.215$\pm$0.005 \cite{son1969velocity}\cite{cardell1993flow}} & \multicolumn{1}{l}{1.33$\pm$0.05\cite{cardell1993flow}} \\
   & \multicolumn{1}{l}{} & \multicolumn{1}{l}{} & \multicolumn{1}{l}{0.21$\pm$ 0.005\cite{ong1996velocity}} & \multicolumn{1}{l}{1.18$\pm$0.05\cite{lourenco1994characteristics} } \\\hline
    \end{tabular}
    \captionsetup{font={footnotesize}}
    \caption{Global flow quantities in the cylinder flow computation at $Re_{D}=3900$. Here $S_t=fD/U_\infty$, $f$ is the main shedding frequency of the vortices, $C_d= \frac{F_x}{{1}{2}\rho U_{\infty}^2 A} $ and $C_l=\frac{F_y}{{1}{2}\rho U_{\infty}^2 A}$ with $A$ standing for the reference area and $F_x$ and $F_y$ representing $x-$ and $y-$components of the total fluid force acting on the cylinder, respectively. $\overline{L_r}$ is the recirculation length, $D$ is the diameter of the cylinder.}
    \label{tab:cylinder2}
\end{table}

\subsubsection{Simulation on coarse mesh}
\begin{figure}[ht!]
    \centering
    \includegraphics[height=0.16\paperheight]{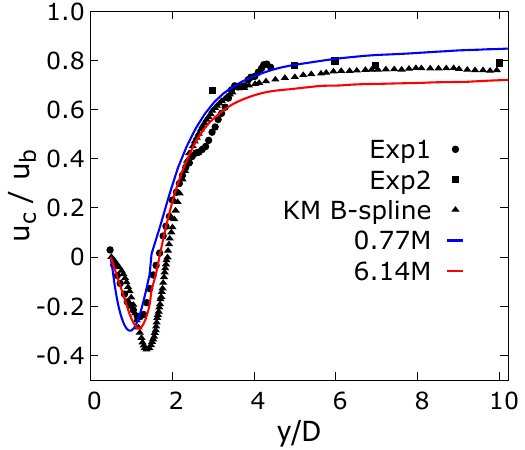}
    \includegraphics[height=0.16\paperheight]{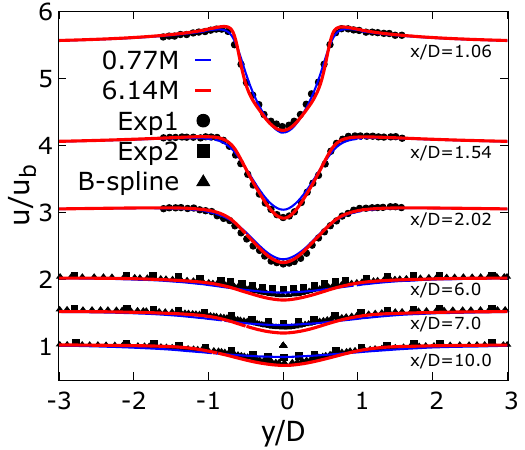}
    \includegraphics[height=0.16\paperheight,width=0.32\linewidth]{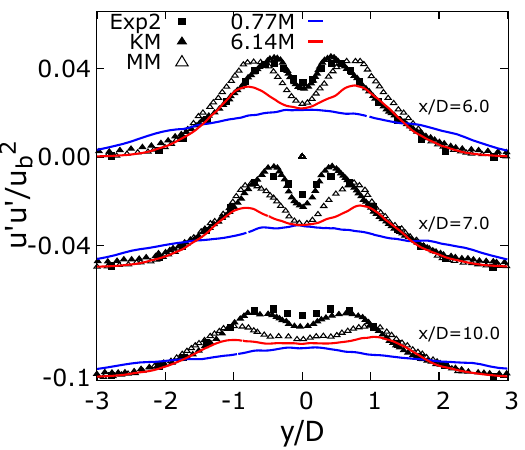}
    \captionsetup{font={footnotesize}}
    \caption{Left: The mean velocity scaled by the bulk velocity in the center line $u_c/u_b$. Middle: the mean streamwise velocity at six different locations. Right: mean streamwise velocity fluctuation $u'u'$ at three downstream locations behind a circular cylinder at $Re_D=3900$. 0.77M: the QR simulation with 0.77 million mesh points (Run \RomanNumeralCaps{1})($\textcolor{blue}{-}$); 6.14M: the QR simulation with 6.14 million mesh points (Run \RomanNumeralCaps{2})(\textcolor{red}{$-$}). Reference data: the experiment of Lourenco and Shih ($\bullet$), the experiment from Ong and Wallace($\blacksquare$), the B-spline large eddy simulation of Kravchenko and Moin($\blacktriangle$), the central difference LES simulation from Mittal and Moin($\triangle$), the upwind LES simulation of Beaudan and Moin($\lozenge$), central difference LES simulation of Breuer($\circ$).}
    \label{fig:cylinder2}
\end{figure}

The first simulation (Run \RomanNumeralCaps{1}) was carried out on a coarse mesh with about 0.77 million grid points. The results from this simulation are shown in Figure \ref{fig:cylinder2}.  The profiles of mean streamwise velocity in the downstream region are in agreement with the experiments of Ong and Wallace. The Reynolds stresses $u'v'$, however, are significantly higher than those of the reference LES and experiments of Ong and Wallace. Furthermore, the recirculation region is smaller and no fluctuation in streamwise velocity was observed. This indicates that the flow has not developed enough three-dimensionality, because the mesh resolution is inadequate to resolve the separating shear layer. Therefore, it was decided to continue the simulation on a mesh with increased resolution. 

\subsubsection{Increasing mesh resolution}
In the simulation Run \RomanNumeralCaps{2}, with the domain consisting of $6.1$ million grid points, the flow field was interpolated from the coarser mesh and advanced in time for approximately twenty shedding cycles ($T=100D/U_\infty$) to allow all the transients to exit the computational domain.
The overall agreement of the streamwise velocity resulting from the filtered central finite difference (Run \RomanNumeralCaps{2}) and the experimental results of Lourenco and Shih is excellent at the first three locations $x/D=1.06$, $1.54$ and $2.02$.  The simulation presents V-shape profiles for the streamwise velocity in the recirculation region at $x/D=1.06$ and $1.54$. The range of the recirculation region $0.5\le  x/D \le 1.69$ predicted by this simulation is noticeably consistent with the experiment. The streamwise Reynolds stress $u'u'$ in the very near wake region $x/D=1.54$ (we do not show the figure here) is quite well predicted in comparison with measurements by Lourenco and Shih and Breuer's simulation which adopted Smagorinsky sub-grid model. 

However, there are some significant deviations between the results obtained from the QR simulation and those of the B-spline simulation and the experiment of Ong and Wallace. In the downstream location, the filtered central difference method (Run \RomanNumeralCaps{2}) underestimates the streamwise mean velocity and the peak of velocity fluctuations $u'u'$ at $x/D = 6, 7$ and $10$ and displays slightly low levels of Reynolds shear stress $u'v'$ at $x/D = 6$ and $7$. The shape of the mean velocity profile is directly related to the level of velocity fluctuations and, consequently, to the transition in the shear layers.

\subsubsection{The influence of discretization scheme}

\begin{figure}[ht!]
\begin{subfigure}[b]{0.5\textwidth}
    \centering
    \captionsetup{font={footnotesize}}
    \includegraphics[width=1\textwidth]{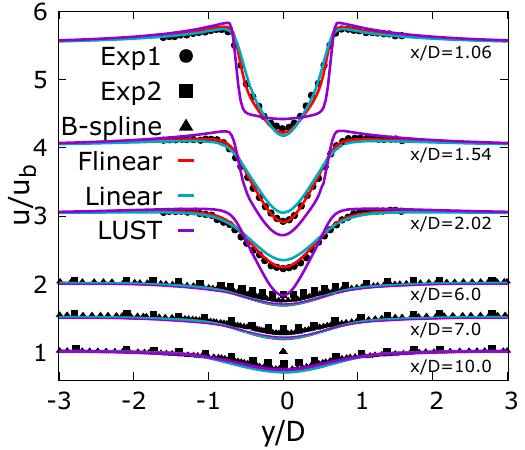}
    \caption{The mean velocity at six positions}
    \includegraphics[width=1\textwidth]{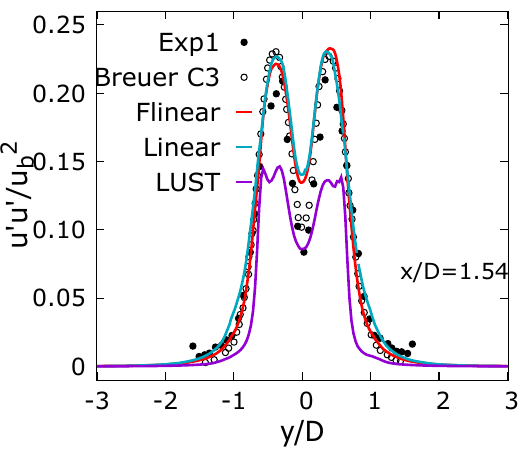}
    \caption{The streamwise Reynolds stress at $x/D=1.54$}
\end{subfigure}
\begin{subfigure}[b]{0.5\textwidth}
    \centering
    \includegraphics[width=1\textwidth]{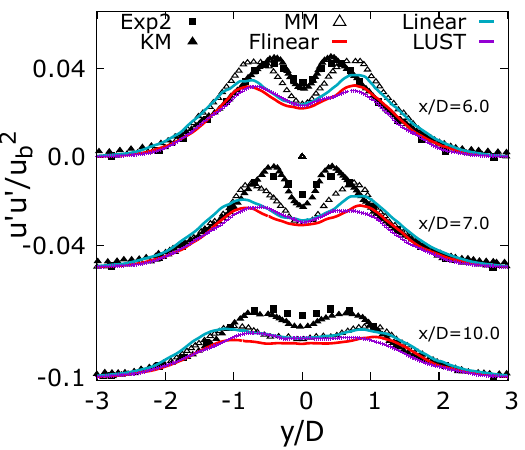}
    \captionsetup{font={footnotesize}}
    \caption{The Reynolds stress at $x/D=6, 7$ and $ 10$}
    \includegraphics[width=1\textwidth]{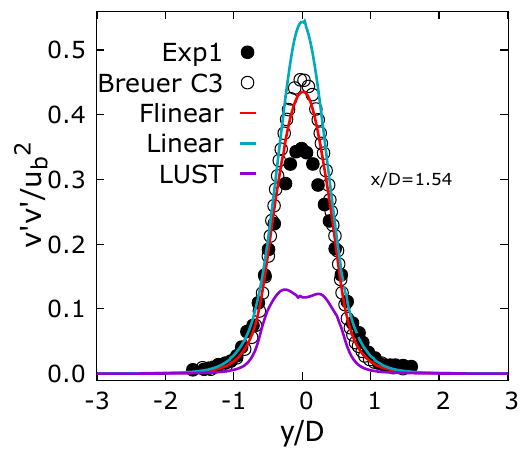}
    \captionsetup{font={footnotesize}}
    \caption{The Reynolds stress in cross-section at $x/D=1.54$}
\end{subfigure}
\captionsetup{font=small}
\caption{The flow over cylinder at $Re_D=3900$ with domain size of $50D \times 30D \times \pi D$. Flinear: QR simulation with a filtered central difference(Run \RomanNumeralCaps{2})($\textcolor{blue}{-}$); Linear: QR simulation with a pure central difference(Run \RomanNumeralCaps{3})($\textcolor{blue}{-}$); LUST: QR simulation with upwind-biased(Run \RomanNumeralCaps{4})(\textcolor{purple}{$-$}).  For details of reference data, see the caption for Figure \ref{fig:cylinder2}.}
\label{fig:cylinder3}
\end{figure}

To study the influence of the finite volume discretization methods for the convective fluxes and the pressure gradient on the turbulence behaviors, simulations with $100\%$ central difference schemes (Run \RomanNumeralCaps{3}) and $25\%$ upwind-biased central difference (Run \RomanNumeralCaps{4}) methods are performed at the same mesh resolution of the filtered central difference (Run \RomanNumeralCaps{2}), i.e. 6.14 million points.

Figure \ref{fig:cylinder3} shows that the mean streamwise velocity $u$ obtained from the three numerical schemes differs in the near wake region ($x/D < 2.02$). The upwind-blended scheme (Run \RomanNumeralCaps{4}), predicts a U-shape mean velocity at $x/D = 1.06$, and develops a much lower V-shape profile at $x/D = 1.54$ and $ 2.02$, compared to other two schemes. 
The central difference simulation highlights that the transition to turbulence in the separating shear layers occurs closer to the cylinder and leads to the development of the V-shape profile and shorter vortex formation region. As a result, the shear layers are shorter and the recirculation region is smaller. 
In the downstream location, it is found that varying the numeric schemes has no apparent effect on the mean velocity. However, the periodic hill simulations demonstrate that the choice of numerical schemes significantly affects the mean and root-mean-square variables.

For cross-section Reynolds stress $v'v'$ at $x/D=1.54$, the upwind-blended scheme (Run \RomanNumeralCaps{4}) predicts lower values. The pure central difference scheme (Run \RomanNumeralCaps{3}) calculates values that are  too large. In addition, the filtered central difference (Run \RomanNumeralCaps{2}) is in accordance with Breuer's simulation\cite{breuer1998large} which used the Smagorinsky sub-grid model conjugated with a central difference, but both Run \RomanNumeralCaps{2} and Breuer's simulations overestimate $v'v'$ in comparison with the experiments of Lourenco and Shih. 

As for the streamwise Reynolds stress $u'u'$ (shown in Figure \ref{fig:cylinder3}), the central difference (Run \RomanNumeralCaps{3}) improves the velocity fluctuation $u'u'$ to a small extent everywhere in the downstream region.  However, the minor differences between Run \RomanNumeralCaps{2} and Run \RomanNumeralCaps{3} are too small to distinguish one technique from another. 

Figure \ref{fig:cylinder3} depicts the mean velocity at the central line. It is clear that the solutions of the filtered central difference (Run \RomanNumeralCaps{2}) match the experimental data\cite{lourenco1994characteristics} very well in the near wake region. Furthermore, the recirculation length of $1.189$ is in good agreement with the experimental\cite{lourenco1994characteristics} value of $1.18$. The upwind-blended scheme (Run \RomanNumeralCaps{4}), however, calculates a conspicuously long region of recirculation, but then shows good agreement of center line streamwise velocity further downstream, at $x/D > 7$, comparing with the B-spline simulation and the hot-wire measurements of Ong and Wallace. 

Additionally, a pure central difference Run \RomanNumeralCaps{5} with about ten million grid points was carried out. To see if this simulation provides enough resolution in the downstream region ($x/D > 6$) to improve the underprediction of streamwise velocity at the center line and Reynolds stress. The mean velocity appears to be unaffected by increasing the resolution. The simulation (Run \RomanNumeralCaps{5}) calculates larger fluctuations ($u'u'$ and $v'v'$) in the entire flow domain compared with the coarse mesh. The larger fluctuation matches Ong's experimental data in the downstream locations ($x/D >6.0$) but disagrees with Lourenco's results in the near wake region ($x/D=1.54$).  
Note that the two sets of experimental data are inconsistent with each other.

Finally, to study the impact of the size of the periodic domain, Run \RomanNumeralCaps{6} is performed in which the spanwise size is doubled while keeping the mesh resolution as in Run \RomanNumeralCaps{2}.
The mean and Reynolds stresses are compared between the small domain with the length of $\pi D$ and the larger domain with the length of $2 \pi D$ in the spanwise direction. 
There is no evidence that doubling the periodic domain has an influence on mean velocity and Reynolds stress. In conclusion, the comparison of Run \RomanNumeralCaps{6} and Run \RomanNumeralCaps{2} show that the domain size of $50D \times 30D \times \pi D$ is large enough.

\subsubsection{Minimum Resolution}
\begin{figure}[ht!]
    \centering
    \captionsetup{font={footnotesize}}
    \includegraphics[width=0.45\linewidth]{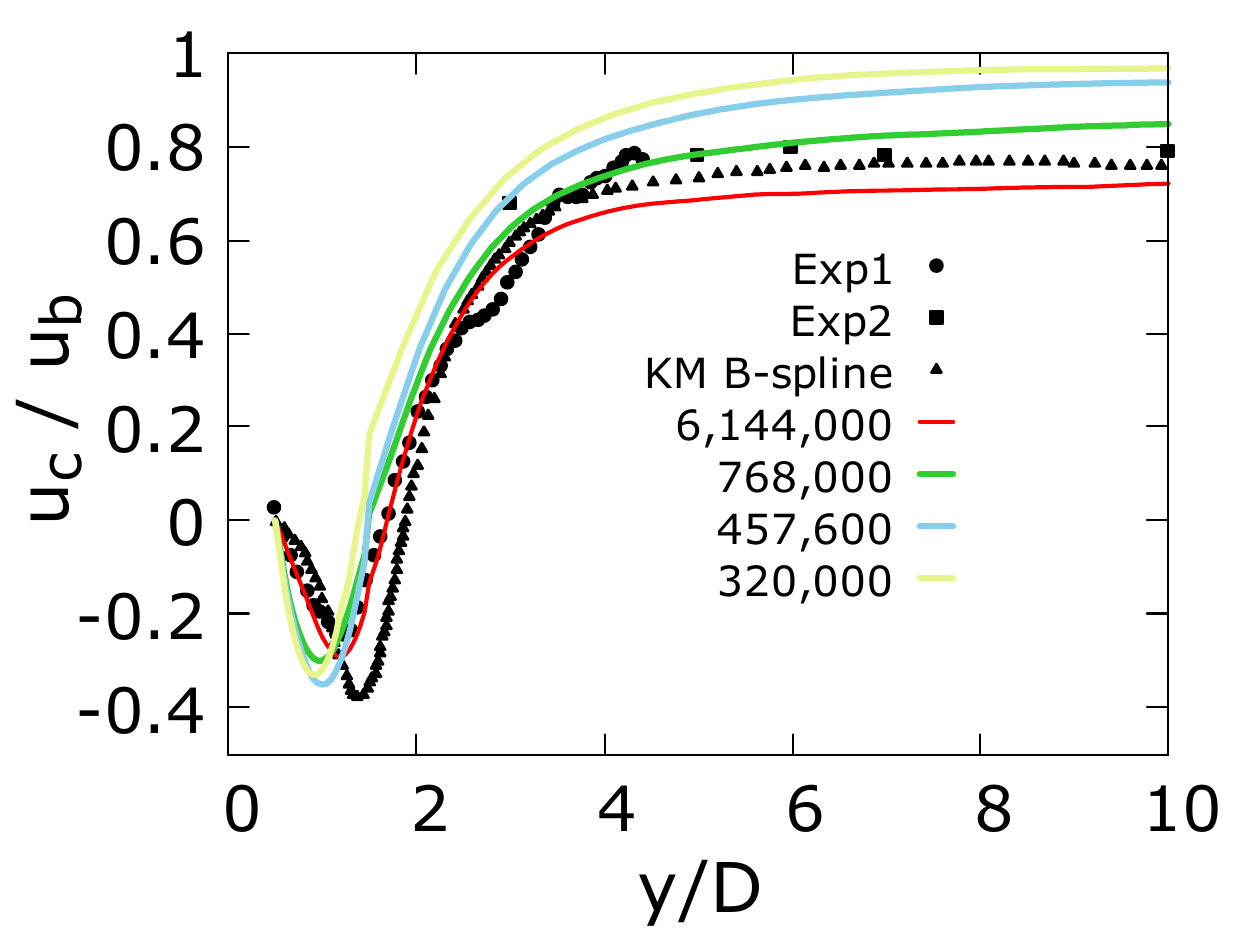}
    \includegraphics[width=0.45\linewidth]{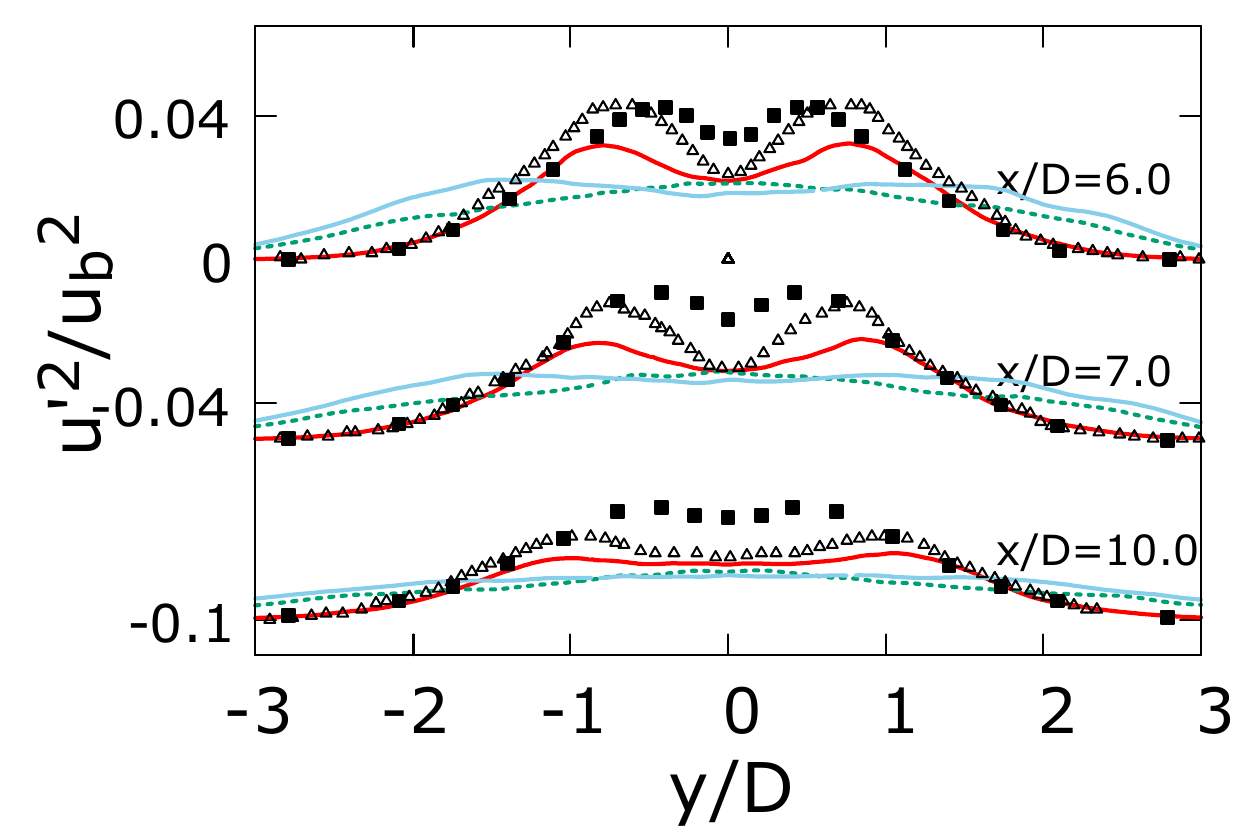}
\captionsetup{font=small}
\caption{The flow over a cylinder at $Re_D=3900$ with different mesh resolutions to find the minimum resolution. Left: the mean streamwise velocity at the central line; right: the streamwise Reynolds stress at $x/D=1.54$.  For details, see the caption for Figure \ref{fig:cylinder2}.}
\label{fig:cylinder6}
\end{figure}
The minimum resolution required to accurately simulate the flow over a cylinder was investigated on four meshes with 768,000, 457,600, 320,000, and 204,800 grid points, respectively. The results of these simulations are presented in Figure \ref{fig:cylinder6}. In the near wake region ($x/D<2.02$), where viscosity has the most significant impact, the QR model accurately predicts the mean streamwise velocity, spanwise velocity, and streamwise Reynolds stress. However, in the downstream region ($x/D>6$), the mesh is stretched resulting in a lower resolution, and thus, no fluctuations were observed. To improve the results in the downstream region, a finer mesh should be used. For those interested in the near wake region where separation and recirculation occur, the coarsest mesh with approximately 0.4 million grid points is sufficient.

\subsubsection{The reason for discrepancy}
The significant difference in the size and formation of the recirculation region directly affects the length of the vortex formation and the dynamics of the downstream flow.
The discrepancy between filtered central difference and reference data in downstream locations ($x/D = 6.0, 7.0, 10$) can be caused by many factors. One of the possible reasons can be attributed to various levels of free-stream turbulence present in different simulations. Gerrard\cite{gerrard1966mechanics} mentioned that the size of the vortex formation region becomes smaller by increasing the freestream turbulence level with $1\%$. 

It has been reported that a shorter recirculation region leads to shorter vortex formation in the downstream region. In the B-spline simulations\cite{kravchenko2000numerical}, the shear layer is larger and the recirculation region is longer, which differs from the experimental data of Lourenco and Shih. Consequently, the development of the flow downstream is different. Therefore, the current simulations are comparable qualitatively, instead of quantitatively, to that of B-spline simulations and experiments of Ong and Wallace in the downstream region.  Additionally, the difference to the experiments has also been attributed to experimental errors as manifested in the large asymmetry of the experimental data\cite{lourenco1994characteristics}.

\subsubsection{Conclusion}
In general, the filtered central difference scheme (Run \RomanNumeralCaps{2}) gives the best results. The consistency of the QR model simulations is illustrated by the fact that the numerical solution of the mean variable approaches each other and the experimental results as the time step and grid spacing tend to zero in the solution domain. 

The simulation results strongly depend on the finite volume discretization methods. It is found that the QR model combined with filtered central difference yields the best prediction of the mean velocity and Reynolds stress. The filtered central difference is in good agreement with the experiments carried out by Lourenco and Shih which were limited to the near wake region ($x/D<2.02$), despite the fact that the filtered central difference discretization does not agree well with the experimental data of Ong and Wallace\cite{ong1996velocity} and B-spline LES of Kravchenko and Moin\cite{kravchenko2000numerical} in the downstream location ($x/D>6$).
The outcomes reveal that increasing the mesh resolution has no effect on the mean velocity but improves the Reynolds stress $u'u'$ in the downstream region. Doubling the domain size in the periodic direction does not affect the results. Finally, the investigation reveals that the minimum resolution for the mean velocity and Reynolds stress in the near wake region is about 0.4 million points.

\section{Conclusion}
A thorough comparison between the minimum-dissipation model (QR) of large eddy simulations (LES) and experimental/numerical results for channel flow, flow past a circular cylinder, and flow over periodic hills shows generally favorable agreement.

In channel flow, the results indicate that the static QR model performs equally well as dynamic models while reducing the computational cost. The model constant of $C=0.024$ yields the most accurate predictions, and the contribution of the sub-grid model diminishes as mesh resolution increases, becoming small (less than 0.2 times the molecular viscosity) at the finest mesh. Moreover, the QR model accurately predicts turbulence mean and invariance up to $Re_\tau = 2000$ using relatively coarse meshes and without using any wall damping function. At $Re_\tau = 1000$, the symmetry-preserving discretization performs better than the standard OpenFOAM discretization.

For flow over periodic hills, various comparisons demonstrate the necessity of using symmetry-preserving discretization or central difference schemes in OpenFOAM alongside the minimum dissipation model. By increasing the Reynolds number by approximately $3\%$ to $7\%$, The results agree better with the reference data. The model constant of $C=0.024$ is again the best choice.

Regarding flow over a cylinder, the mean velocity, drag coefficient, and lift coefficient exhibit good agreement with experimental data. The behavior of turbulence strongly relies on the finite volume discretization methods. The combination of the QR model with filtered central difference yields the best approximation of the mean velocity and Reynolds stress. Increasing mesh resolution has minimal effect on the mean variable but improves the stress $u'u'$ in the downstream region. Doubling the domain size in the periodic direction does not impact the results. Finally, the investigation shows that a minimum resolution of approximately 0.4 million is necessary for an accurate representation of mean velocity and Reynolds stress in the near wake region.

\begin{center}
\bibliographystyle{plain} 
\bibliography{refs} 
\end{center}

\end{document}